\documentclass[lettersize,journal]{IEEEtran}

\UseRawInputEncoding
\usepackage{amssymb}
\usepackage[table,xcdraw]{xcolor}
\usepackage{bm}
\usepackage{color}
\usepackage{graphicx, floatrow}
\usepackage{stfloats}  
\usepackage{float}  
\floatstyle{plaintop}
\restylefloat{table}
\usepackage{empheq}
\usepackage[linesnumbered,ruled,lined]{algorithm2e}
\usepackage[outdir=./]{epstopdf}
\usepackage{lipsum}
\usepackage{cite}
\usepackage{multirow}
\usepackage{hyperref}
\hypersetup{hidelinks,
	colorlinks=true,
	allcolors=black,
	pdfstartview=Fit,
	breaklinks=true}
\usepackage{subcaption}
\usepackage{fancyhdr}
\SetKwInput{Input}{input}
\SetKwInput{Output}{output}

\usepackage{array}
\newcolumntype{L}[1]{>{\raggedright\let\newline\\\arraybackslash\hspace{0pt}}m{#1}}
\newcolumntype{C}[1]{>{\centering\let\newline\\\arraybackslash\hspace{0pt}}m{#1}}
\newcolumntype{R}[1]{>{\raggedleft\let\newline\\\arraybackslash\hspace{0pt}}m{#1}}
\usepackage{amsmath}
\DeclareMathOperator{\tr}{tr}

\usepackage{booktabs}
\newcolumntype{I}{!{\vrule width 1.05pt}}

\IEEEoverridecommandlockouts
\usepackage[acronym]{glossaries-extra}
\setabbreviationstyle[acronym]{long-short}
\newacronym{aod}{AoD}{angle of departure}
\newacronym{ao}{AO}{alternating optimization}
\newacronym{asr}{ASR}{average sum-rate}
\newacronym{awgn}{AWGN}{additive white gaussian noise}
\newacronym{awmse}{AWMSE}{augmented weighted MSE}
\newacronym{csir}{CSIR}{channel state information at the receiver}
\newacronym{csit}{CSIT}{channel state information at the transmitter}
\newacronym{csi}{CSI}{channel state information}
\newacronym{dd}{DD}{delay-Doppler}
\newacronym{dft}{DFT}{discrete Fourier transform}
\newacronym{esr}{ESR}{ergodic sum-rate}
\newacronym{fft}{FFT}{fast Fourier transform}
\newacronym{ifft}{IFFT}{inverse fast Fourier transform}
\newacronym{ift}{IFT}{inverse Fourier transform}
\newacronym{iid}{i.i.d}{independent and identically distributed}
\newacronym{isfft}{ISFFT}{inverse symplectic finite Fourier transform}
\newacronym{kkt}{KKT}{Karush-Kuhn-Tucker}
\newacronym{leo}{LEO}{Low Earth Orbit}
\newacronym{lln}{LLN}{law of large numbers}
\newacronym{los}{LoS}{line of sight}
\newacronym{mimo}{MIMO}{multiple-input multiple-output}
\newacronym{mmse}{MMSE}{minimum mean square error}
\newacronym{mse}{MSE}{mean square error}
\newacronym{nlos}{NLoS}{non-line of sight}
\newacronym{noma}{NOMA}{non-orthogonal multiple access}
\newacronym{ofdm}{OFDM}{orthogonal frequency division multiplexing}
\newacronym{otfs}{OTFS}{Orthogonal time frequency space}
\newacronym{qcqp}{QCQP}{quadratically constrained quadratic program}
\newacronym{rsma}{RSMA}{rate-splitting multiple access}
\newacronym{rs}{RS}{rate-splitting}
\newacronym{saa}{SAA}{sample average approximation}
\newacronym{saf}{SAFs}{sample average functions}
\newacronym{sdma}{SDMA}{spatial division multiple access}
\newacronym{sfft}{SFFT}{symplectic finite Fourier transform}
\newacronym{sic}{SIC}{successive interference cancellation}
\newacronym{sinr}{SINR}{signal to interference-plus-noise ratio}
\newacronym{siso}{SISO}{single-input single-output}
\newacronym{snr}{SNR}{signal-to-noise ratio}
\newacronym{td}{TD}{time-delay}
\newacronym{tf}{TF}{time-frequency}
\newacronym{tsl}{TSL}{terrestrial-satellite link}
\newacronym{ula}{ULA}{uniform linear array}
\newacronym{wmmse}{WMMSE}{weighted MMSE}
\newacronym{ber}{BER}{bit error rate}
\newacronym{zf}{ZF}{zero-forcing}
\newacronym{dof}{DoF}{degree-of-freedom}
\newacronym{wmse}{WMSE}{weighted MSE}

\begin{document}
\title{Robust Design for Multi-Antenna LEO Satellite Communications with Fractional Delay and Doppler Shifts: An RSMA-OTFS Approach}

\author{
    \IEEEauthorblockN{Yunnuo~Xu, \IEEEmembership{Member, IEEE}, Yumeng Zhang, Yijie~Mao, \IEEEmembership{Member, IEEE}, \\Bruno~Clerckx, \IEEEmembership{Fellow, IEEE},  Yun Hee~Kim, \IEEEmembership{Senior Member, IEEE}, Yujun Li, \IEEEmembership{Member, IEEE}}
    \thanks{
        \par Y. Xu and Y. Li are with the School of Information Science and  Engineering, Shandong University, Qingdao 266237, China (email: \{yunnuo.xu, liyujun\}@sdu.edu.cn).
        \par Y. Zhang is with the Department of Electronics and Computer Engineering, The Hong Kong University of Science and Technology, Hong Kong 999077, China (e-mail: eeyzhang@ust.hk).
        \par Y. Mao is with the School of Information Science and Technology, ShanghaiTech University, Shanghai 201210, China (email: maoyj@shanghaitech.edu.cn).
        \par B. Clerckx is with the Department of Electrical and Electronic Engineering at Imperial College London, London SW7 2AZ, UK (email: b.clerckx@imperial.ac.uk).
        \par  Y.H. Kim is with the Department of Electronic Engineering and jointly affiliated with the Electronics and Information Convergence Engineering, Kyung Hee University, Yongin 17104, Korea (e-mail: yheekim@khu.ac.kr).
    }
}

\maketitle

\begin{abstract}
    Low-Earth-orbit (LEO) satellite communication systems face challenges due to high satellite mobility, which hinders the reliable acquisition of instantaneous channel state information at the transmitter (CSIT) and subsequently degrades multi-user transmission performance. This paper investigates a downlink multi-user multi-antenna system, and tackles the above challenges by introducing orthogonal time frequency space (OTFS) modulation and rate-splitting multiple access (RSMA) transmission. Specifically, OTFS enables stable characterization of time-varying channels by representing them in the delay-Doppler domain. However, realistic propagation introduces various inter-symbol and inter-user interference due to non-orthogonal yet practical rectangular pulse shaping, fractional delays, Doppler shifts, and imperfect (statistical) CSIT. In this context, RSMA offers promising robustness for interference mitigation and CSIT imperfections, and hence is integrated with OTFS to provide a comprehensive solution. A compact cross-domain input-output relationship for RSMA-OTFS is established, and an ergodic sum-rate maximization problem is formulated and solved using a weighted minimum mean-square-error based alternating optimization algorithm that does not depend on channel sparsity. Simulation results reveal that the considered practical propagation effects significantly degrade performance if unaddressed. Furthermore, the RSMA-OTFS scheme demonstrates improved ergodic sum-rate and robustness against CSIT uncertainty across various user deployments and CSIT qualities.
\end{abstract}


\section{Introduction}
\par The rapid deployment of \gls{leo} satellite constellations is reshaping the global communication landscape by enabling broadband access in remote and underserved areas and supporting the development of space-air-ground integrated networks \cite{38821}. Compared with terrestrial communications, \gls{leo} satellites operate at low altitudes (typically between $500$ km and $2000$ km above the Earth), which enables wide-area coverage. However, the high orbital velocity induces high-dynamic time-varying channels, resulting in time and frequency dispersion. These impairments introduce inter-symbol interference, hinder accurate channel estimation, and challenge coherent demodulation \cite{9785832,9849120}. In multi-antenna multi-user systems, such rapidly time-varying channels combined with imperfect \gls{csit} further reduce spatial multiplexing efficiency, complicate precoder design, and impose stringent constraints on interference mitigation \cite{guo2025prompt}. Consequently, system performance degradation caused by these practical impairments cannot be overlooked in \gls{leo} communication scenarios \cite{9849120,10283795}.

\par Addressing these challenges requires transmission strategies capable of handling rapid channel variations and mitigating the effects of \gls{csit} uncertainty. \gls{otfs} modulation has emerged as a promising solution for doubly selective channels due to its comprehensive \gls{dd} domain representation and its ability to exploit full \gls{tf} diversity \cite{7925924}. In parallel, \gls{rsma} serves as an effective multiple access technique in managing multi-user interference and enhancing robustness to imperfect \gls{csit} in multi-user multi-antenna networks \cite{9831440}. The complementary capabilities of the two techniques make their integration an appealing approach for addressing the rapidly time-varying characteristics of \gls{leo} system.

\par Research on \gls{otfs} has advanced rapidly in recent years. By converting the time-varying channel in the \gls{tf} domain into a nearly time-invariant structure in the \gls{dd} domain, \gls{otfs} has demonstrated clear benefits over \gls{ofdm} in high-mobility scenarios \cite{7925924,8424569}. The fundamental input-output relationship under \gls{siso} system and a message-passing detection framework were developed in \cite{8424569}, and subsequent refinements exploited \gls{dd} domain sparsity to reduce factor-graph iterations while maintaining competitive \gls{ber} performance \cite{9492800}. Further work includes cross-domain detectors that leverage time domain sparsity and \gls{dd} domain symbol constellation constraints \cite{9536449}, as well as optimized power allocation strategies given frequency-domain \gls{zf} and \gls{mmse} equalizers \cite{9864651}.

\par Beyond symbol detection, \gls{mimo}-\gls{otfs} has also attracted significant attention for further performance enhancement. Tomlinson-Harashima precoding was introduced in \cite{10042436} under sparse-channel assumptions, \cite{9362336} proposed a \gls{dd} domain precoder based on Hermitian channel structure. More recent contributions such as \cite{du2025signal} optimized the signal design to exploit \gls{otfs}'s potential in dual-functional radar-communication, and \cite{10859263} developed low-complexity precoding schemes built on simplified channel models.
Despite these advancements, most existing studies rely on idealized settings in which the \gls{otfs} frame is assumed to be sufficiently large to provide high \gls{dd} resolution, thereby approximating the continuous delay and Doppler shifts by their integer-valued counterparts.
In practical systems, however, fractional Doppler shifts and fractional delays are non-negligible, and together with the use of practical rectangular pulse shaping, they jointly break the canonical doubly block-circulant or sparse structures commonly assumed for \gls{otfs} channel matrices \cite{10283795}. This structural distortion fundamentally limits the applicability of many existing precoding and detection approaches.

\par Besides the modulation scheme, \gls{rsma} is a reliable multiple access strategy to incorporate with \gls{otfs} in \gls{leo} scenarios, due to its resilience against imperfect \gls{csit} and multi-user interference.
Its robustness originates from both its signal structure and its theoretical \gls{dof} optimality under imperfect \gls{csit}. At the transmitter, \gls{rsma} linearly precodes rate-splitted messages, where each user message is divided into a common part decoded by all users and a private part decoded only by its intended user \cite{9831440,9451194,8907421}. This structure enables each receiver to successively remove part of the multi-user interference and to reliably decode the remaining private information even when \gls{csit} is inaccurate. From an information-theoretic perspective, \gls{rsma} has been proven to achieve the optimal \gls{dof} region in a wide range of user configurations with partial \gls{csit}, whereas conventional \gls{sdma} and \gls{noma} are suboptimal in the same settings. These properties lead to substantial gains in system throughput and reliability, together with improved energy efficiency and reduced transmission latency \cite{9663192,9491092,10741240,9831048}.
Its robustness to imperfect \gls{csit} makes it particularly suitable for rapidly time-varying environments, with notable benefits demonstrated in satellite communications  \cite{10896843,10521807}. When combined with \gls{otfs}, \gls{rsma} further strengthens the interference management capability and mitigates \gls{csit}-related performance degradation, both of which are critical in practical \gls{leo} systems.

\par Unlike \gls{ofdm}-based \gls{rsma}, applying \gls{rsma} to \gls{otfs} inherently requires operating in the \gls{dd} domain, where each symbol spreads across the entire two-dimensional \gls{tf} plane. This spreading fundamentally changes how precoders interact with data symbols and complicates the design of transmit strategies. Consequently, new precoding and interference management approach is required to accommodate the two-dimensional channel structure of \gls{otfs}, rather than relying on the per-subcarrier resource allocation used in \gls{ofdm}~\cite{10236464}.
Several studies have examined the integration of \gls{rsma} and \gls{otfs}. The work in \cite{10462183} investigated a cross-domain uplink design that exploits both \gls{tf} and \gls{dd} representations and demonstrated gains in outage performance and user fairness. A downlink extension was presented in \cite{10855588}. Both works, however, rely on ideal channel models that assume block-circulant structure under bi-orthogonal pulse shaping. Practical rectangular pulse shaping was considered in \cite{10804646}, where fractional Doppler effects were included but fractional delays were omitted. Furthermore, all these studies assume perfect \gls{csit}, which is difficult to obtain in \gls{leo} satellite systems due to high mobility and large feedback latency. Most existing work also restricts attention to \gls{siso} configurations without precoder optimization, limiting their ability to fully exploit the benefits of multi-antenna configurations and \gls{rsma}. As a result, multi-antenna \gls{rsma}-\gls{otfs} transmission under realistic channel impairments remains insufficiently explored.

\par Motivated by 1) \gls{otfs} and \gls{rsma} offer complementary benefits in \gls{leo} systems, where \gls{otfs} provides resilience to fast channel dynamics, and \gls{rsma} improves robustness to interference and \gls{csit} uncertainty, 2) existing studies lack a multi-antenna precoding strategy that comprehensively accounts for the imperfect propagation effect in \gls{otfs} \gls{leo} networks, this paper develops a  \gls{rsma}-\gls{otfs} transmission framework that considers realistic propagation with fractional delay and Doppler effects, practical non-orthogonal pulse shaping, and imperfect CSIT. The paper, thereby, aims to reduce the gap between theoretical and practical applications in \gls{leo} systems.
The main contributions of this article are summarized as follows:

\begin{enumerate}
    \item We develop a robust multi-user multi-antenna \gls{rsma}-\gls{otfs} transmission framework suitable for \gls{leo} satellite communications. This framework incorporates practical propagation modelling, including fractional Doppler shifts and delays, practical pulse shaping, and statistical \gls{csit}. It hence extends prior studies that rely on idealized assumptions or \gls{siso} settings and establishes a comprehensive study for robust precoder design under \gls{csit} uncertainty.

    \item Building on the proposed transmission framework, we derive an equivalent cross-domain input-output relationship for multi-antenna \gls{rsma}-\gls{otfs} systems with statistical \gls{csit}. On this basis, we formulate a sum-rate maximization problem that jointly optimizes the multi-antenna precoders, message splitting, power allocation, and the arrangement matrices that map the precoders into the \gls{dd} domain. The proposed formulation captures the interaction between common and private message transmission, and enables robust precoder design in the two-dimensional \gls{dd} domain. To the best of our knowledge, this is the first work to investigate joint precoding and arrangement optimization for multi-antenna \gls{rsma}-\gls{otfs} transmission under statistical \gls{csit} in \gls{leo} satellite systems.

    \item  Due to the non-convex nature of the formulated problem, we transform it into a tractable form through \gls{saa} and vectorization, and employ a \gls{wmmse}-based \gls{ao} algorithm for its solution. The proposed optimization framework does not rely on a sparse representation of the channel in the \gls{dd} domain, which enhances its applicability to practical channel assumptions where such sparsity may not hold. Numerical results demonstrate that the multi-antenna \gls{rsma}-\gls{otfs} design achieves higher \gls{esr} than existing schemes and provides strong robustness to channel uncertainty and statistical \gls{csit} inaccuracy.
\end{enumerate}

\par \textit{Organization:} The remainder of this paper is organized as follows.
Section~\ref{sec:system_model} introduces the proposed \gls{rsma}-\gls{otfs} transmission framework and establishes the corresponding cross-domain input-output relationship.
Section~\ref{sec:problem_formulation} formulates the \gls{esr} maximization problem, and Section~\ref{sec:proposed_algorithm} presents the \gls{wmmse}-based \gls{ao} algorithm developed to address the problem.
Section~\ref{sec:results} provides the numerical results and performance evaluation.
Finally, Section~\ref{sec:conclusion} concludes the paper.

\par \textit{Notations:} Matrices, column vectors, and scalars are denoted by boldface uppercase, boldface lowercase, and standard letters, respectively (e.g., $\mathbf{A}$, $\mathbf{a}$, $a$).
$[\mathbf{A}]_{ij}$ is the $(i,j)$-th element of matrix $\mathbf{A}$.
$\mathbf{I}_a$ stands for the identity matrix of size $a\times a$.
The operators $(\cdot)^T$ and $(\cdot)^H$ represent transpose and conjugate transpose, respectively.
The sets of real and complex numbers are denoted by $\mathbb{R}$ and $\mathbb{C}$, respectively.
The absolute value of a scalar is written as $|\cdot|$, and $\|\cdot\|$ denotes the Euclidean norm.
The operator $\mathrm{tr}(\cdot)$ denotes the trace, $\mathrm{vec}(\cdot)$ is the vectorization operator, and $\mathrm{vec}^{-1}(\cdot)$ reshapes a vector into a matrix of appropriate dimension.
The Kronecker product is written as $\otimes$.
Expectation is represented by $\mathbb{E}\{\cdot\}$, and $\mathrm{Re}(\cdot)$ denotes the real part of a complex quantity.
The notation $\mathcal{CN}(\boldsymbol{\zeta},\boldsymbol{\sigma}^2)$ represents a circularly symmetric complex Gaussian random vector with mean $\boldsymbol{\zeta}$ and covariance matrix $\boldsymbol{\sigma}^2$.

\section{System model} \label{sec:system_model}
\begin{figure*}[t!]
    \centering
    \includegraphics[scale=0.63]{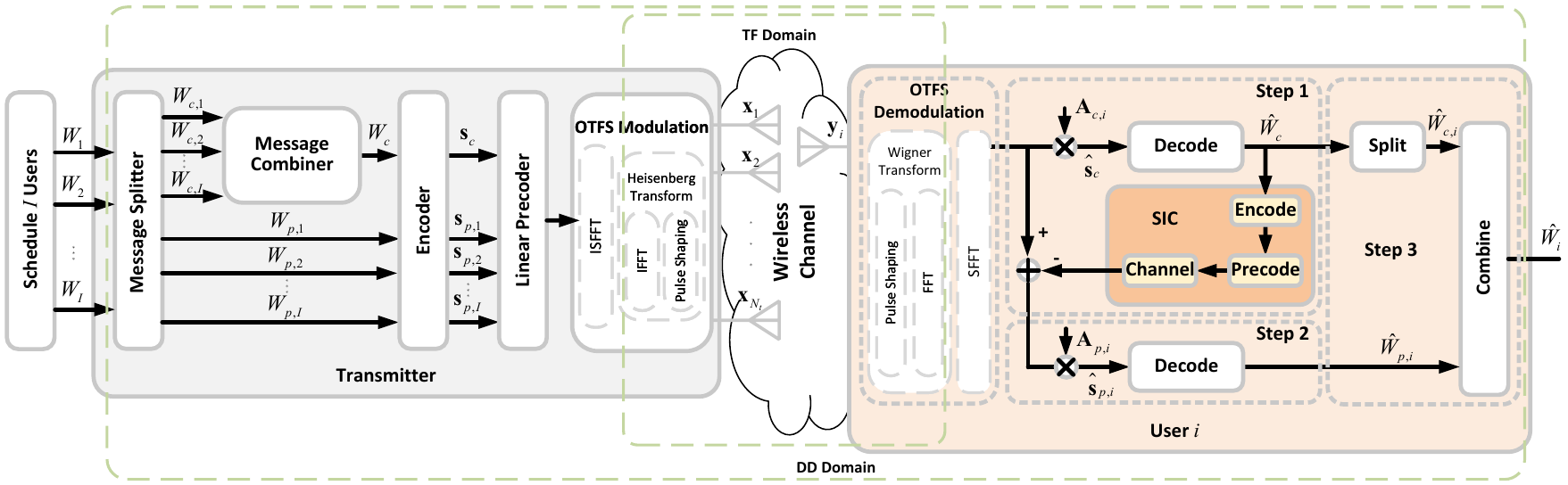}
    \caption{Transmission model of \gls{rsma}-\gls{otfs} in multi-user multi-antenna downlink communications.}
    \label{fig:SystemModel}
\end{figure*}

In this section, we present the system model of a high-speed \gls{leo} satellite equipped with an $N_t$-antenna \gls{ula} serving $I$ single-antenna users simultaneously. The \gls{rsma}-\gls{otfs} transmission scheme is described below in detail. Fig.~\ref{fig:SystemModel} illustrates the general architecture of the multi-user, multi-antenna \gls{rsma}-\gls{otfs} system.

\subsection{Transmitted Signal}\label{subsec:TxSignal}
The information symbols are arranged over \gls{dd} domain with $N$ Doppler bins and $M$ delay bins. We denote $\Delta f$ and $T={1}/{\Delta f}$ as the subcarrier spacing and the time slot duration, respectively.
The \gls{otfs} signal frame is set to $T_f=NT$~s duration and $B=M\Delta f$~Hz bandwidth.
\par
\begin{figure}[t!]
    \centering
    \includegraphics[scale=0.7]{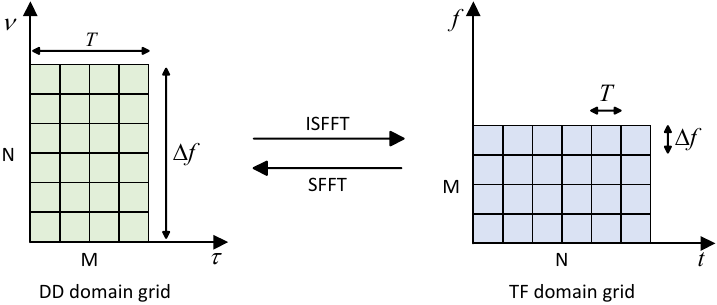}
    \caption{\gls{dd} domain and \gls{tf} domain transformation.}
    \label{fig:TF-DD-transformation}
\end{figure}

 According to \gls{rsma}\footnote{In this study, we adopt 1-layer \gls{rsma} framework, where all common parts are combined into one common message \cite{7555358,Mao2018}.}, each user's message is split into common and private parts, $W_{c,i}$, $W_{p,i}$, $\forall\, i\in\mathcal{I}=\{1,2,\ldots,I\}$. The common parts of all users' messages are combined into a single common message and encoded into a common stream, arranged in \gls{dd} domain $\mathbf{S}_c\in\mathbb{C}^{M\times N}$. The private message of user-$i$ is encoded into private stream arranged in \gls{dd} domain $\mathbf{S}_{p,i}\in\mathbb{C}^{M\times N}$. Correspondingly, we define $\mathbf{s}_c=\text{vec}(\mathbf{S}_c)\in\mathbb{C}^{MN\times1}$, and $\mathbf{s}_{p,i}=\text{vec}(\mathbf{S}_{p,i})\allowbreak\in\mathbb{C}^{MN\times1}$, $\forall\, i\in\mathcal{I}$. 
We assume that the symbol power is normalized. The common and private symbols for all users are first precoded in the DD domain symbols as follows
\begin{equation}
    \mathbf{X}^\mathrm{DD} = \boldsymbol{\Psi}_c\mathbf{s}_c\mathbf{p}_c^T+\sum_{i=1}^{I}\boldsymbol{\Psi}_{p,i}\mathbf{s}_{p,i}\mathbf{p}_{p,i}^T,
\end{equation}
where $\boldsymbol{\Psi}_c\in\mathbb{R}^{MN\times MN}$ and $\boldsymbol{\Psi}_{p,i}\in\mathbb{R}^{MN\times MN}$ are diagonal binary arrangement matrices that map the symbols into the corresponding positions in $\mathbf{X}^\mathrm{DD}$,
and $\mathbf{p}_c$ and $\mathbf{p}_{p,i}\in\mathbb{C}^{N_t\times1}$ are the precoding vectors for common and private symbols, respectively. Denote $\mathbf{P}_{BF}=[\mathbf{p}_c, \allowbreak \mathbf{p}_{p,1},\allowbreak \cdots, \allowbreak\mathbf{p}_{p,I}]^T$ and $\boldsymbol{\Psi}=\{\boldsymbol{\Psi}_c, \allowbreak \boldsymbol{\Psi}_{p,1}, \allowbreak \cdots, \allowbreak \boldsymbol{\Psi}_{p,I}\}$. Then, the vectorized \gls{dd} domain transmitted symbols for all antennas is expressed as
\begin{equation}
    \begin{aligned}
        \mathbf{x}^\mathrm{DD} & =\text{vec}(\mathbf{X}^\mathrm{DD})                                                                                                  \\
                               & =(\mathbf{p}_c\otimes\boldsymbol{\Psi}_c)\mathbf{s}_c+ \sum_{i=1}^{I}(\mathbf{p}_{p,i}\otimes\boldsymbol{\Psi}_{p,i})\mathbf{s}_{p,i}.
    \end{aligned}
\end{equation}
The transmit power is constrained by $P_\mathrm{t}$, i.e., $\mathbb{E}\{\mathbf{P}_{BF}\mathbf{P}_{BF}^H\}\leq P_\mathrm{t}$.
Then \gls{dd} domain symbols $\mathbf{X}^{\mathrm{DD}}$ are sequentially transformed into \gls{tf} and \gls{td} symbols, denoted as $\mathbf{X}^{\mathrm{TF}}$ and $\mathbf{X}^{\mathrm{TD}}$, through \gls{isfft} and the Heisenberg transform, respectively \cite{10042436,7925924}. This process can be expressed as
\begin{subequations}
    \begin{align}
        \mathbf{X}^\mathrm{TF} & = \left(\mathbf{F}_N^H\otimes\mathbf{F}_M\right)\mathbf{X}^\mathrm{DD},                                                                                                       \\
        \mathbf{X}^\mathrm{TD} & =\mathbf{G}_{tx}\left[\left(\mathbf{I}_N\otimes\mathbf{F}_M^H\right)\mathbf{X}^\mathrm{TF}\right]   =\left(\mathbf{F}_N^H\otimes \mathbf{I}_{M}\right)\mathbf{X}^{DD},
    \end{align}
\end{subequations}
where $\mathbf{G}_{tx}=\mathbf{I}_{MN}$ for the rectangular shape \cite{8516353} while $\mathbf{F}_M\in\mathbb{C}^{M\times M}$ and $\mathbf{F}_N\in\mathbb{C}^{N\times N}$ are normalized \gls{dft} matrices. The transformation between the \gls{dd} domain and the \gls{tf} domain is shown in Fig. \ref{fig:TF-DD-transformation}.
The vectorized transmitted symbols in the \gls{td} domain  can be defined as
\begin{equation}
    \begin{aligned}
         & \mathbf{x}^\mathrm{TD}  =\text{vec}(\mathbf{X}^\mathrm{TD})                                                                   \\
         & =\left[\mathbf{p}_c\otimes (\mathbf{F}_N^H\otimes \mathbf{I}_{M})\boldsymbol{\Psi}_c\right]\mathbf{s}_c+
        \sum_{i=1}^{I}\left[\mathbf{p}_{p,i}\otimes (\mathbf{F}_N^H\otimes \mathbf{I}_{M})\boldsymbol{\Psi}_{{p,i}}\right]\mathbf{s}_{p,i} \\
         & =(\mathbf{p}_c\otimes \tilde{\boldsymbol{\Psi}}_c)\mathbf{s}_c+
        \sum_{i=1}^{I}(\mathbf{p}_{p,i}\otimes \tilde{\boldsymbol{\Psi}}_{{p,i}})\mathbf{s}_{p,i},
    \end{aligned}
\end{equation}
where $\tilde{\boldsymbol{\Psi}}_c$ and $\tilde{\boldsymbol{\Psi}}_{p,i}$ are denoted as effective arrangement matrices in \gls{td} domain.

\subsection{Received Signal}\label{RxSignal}
At $i$-th user, the received \gls{td} domain signal is expressed as
\begin{equation}
    \mathbf{y}_i^{\mathrm{TD}}  = \mathbf{H}_i^{\mathrm{TD}}\mathbf{x}^{\mathrm{TD}} + \mathbf{n}_i\, \in\mathbb{C}^{MN\times 1}
    \label{eq:y_dd_vec}
\end{equation}
where 
$\mathbf{H}_i^\mathrm{TD}\in\mathbb{C}^{MN\times N_tMN}$ is the \gls{td} domain channel matrix between the satellite and user-$i$, and $\mathbf{n}_i\in\mathbb{C}^{MN\times 1}$ is the \gls{awgn} vector of user $i$ with $\mathbf{n}_i\sim\mathcal{CN}(0,\sigma_{n,i}^2\mathbf{I}_{MN})$.
Then, the received signal is reshaped into the following matrix representation $\mathbf{Y}_i^{\mathrm{TD}} = \mathrm{vec}^{-1}(\mathbf{y}_i^{\mathrm{TD}})$, where \( \mathbf{Y}_i^{\mathrm{TD}} \in \mathbb{C}^{M \times N} \) denotes the received \gls{otfs} frame in the \gls{td} domain.
\par To obtain its representation in the \gls{tf} domain, Wigner transform (filtering pulse followed by a \gls{fft}) is applied, yielding
\begin{equation}
    \mathbf{Y}_i^{\mathrm{TF}} = \mathbf{F}_M \mathbf{G}_{rx} \mathbf{Y}_i^{\mathrm{TD}},
\end{equation}
where \( \mathbf{G}_{rx} \) denotes the pulse-shaping matrix associated with the receiver filter. Similar to $\mathbf{G}_{tx}$, the receiver pulse is taken to be an identity matrix. 

\par Subsequently, the \gls{sfft} is employed to map the \gls{tf} domain signal into the \gls{dd} domain:
\begin{equation}
    \label{eq:y_dd_mx}
    \mathbf{Y}_i^{\mathrm{DD}} = \mathbf{F}_M^{H} \mathbf{Y}_i^{\mathrm{TF}} \mathbf{F}_N = \mathbf{Y}_i^{\mathrm{TD}} \mathbf{F}_N.
\end{equation}

\par In the vectorized form, the input--output relationship in the \gls{dd} domain can be expressed as
\begin{subequations}
    \label{eq:y_dd_vec_all}
    \begin{align}
        \mathbf{y}_i^{\mathrm{DD}}
         & = (\mathbf{F}_N \otimes \mathbf{I}_M) \mathbf{H}_i^\mathrm{TD} \mathbf{x}^{\mathrm{TD}} + \mathbf{n}_i,                                                                                          \\
         & = (\mathbf{F}_N \otimes \mathbf{I}_M) \mathbf{H}_i^{\mathrm{TD}} (\mathbf{p}_c\otimes \tilde{\boldsymbol{\Psi}}_c)\mathbf{s}_c  \notag                                                           \\
         & \quad + \sum_{i=1}^{I}(\mathbf{F}_N \otimes \mathbf{I}_M) \mathbf{H}_i^{\mathrm{TD}}(\mathbf{p}_{p,i}\otimes \tilde{\boldsymbol{\Psi}}_{p,i})\mathbf{s}_{p,i} + \mathbf{n}_i\label{eq:y_dd_full} \\
         & = \tilde{\mathbf{H}}_i^{\mathrm{DD}}\tilde{\mathbf{P}}_{c}\mathbf{s}_c
        +\sum_{i=1}^{I}\tilde{\mathbf{H}}_i^{\mathrm{DD}}\tilde{\mathbf{P}}_{p,i}\mathbf{s}_{p,i}+\mathbf{n}_i,\label{eq:y_dd_eff}
    \end{align}
\end{subequations}
with
\begin{subequations}
    \begin{align}
         & \tilde{\mathbf{H}}_i^{\mathrm{DD}}=(\mathbf{F}_N \otimes \mathbf{I}_M) \mathbf{H}_i^{\mathrm{TD}},                                          \\
         & \tilde{\mathbf{P}}_{j}=\mathbf{p}_{j}\otimes \tilde{\boldsymbol{\Psi}}_j,\,j\in\left\{c,\{p,i\}_{i=1}^{I}\right\}\label{eq:precoder_dd2td}.
    \end{align}
\end{subequations}
$\tilde{\mathbf{H}}_i^{\mathrm{DD}}$ represents the effective \gls{dd} domain channel, and $\tilde{\mathbf{P}}_{j}$ denotes the effective \gls{td} domain precoding matrix. Note that~\eqref{eq:precoder_dd2td} transfers the \gls{dd} domain precoder, $\mathbf{p}_{j}^T$, to effective \gls{td} domain precoder, $\tilde{\mathbf{P}}_{j}$, $j\in\{c,\{p,i\}_{i=1}^{I}\}$.

\par \textit{Remark 1: In this paper, we claim~\eqref{eq:y_dd_eff} as a cross-domain modeling as it features a mixed \gls{td} ($\tilde{\mathbf{P}}_{j}$) and \gls{dd} ($\tilde{\mathbf{H}}_i^{\mathrm{DD}}$) expression, yet enabling  a decoupling relationship between them, which constitutes one of the novelties of the proposed framework. The decoupled precoding and channel matrices provides a more tractable optimization problem as it will be detailed in Section \ref{sec:proposed_algorithm}. In contrast, conventional approaches typically adopt a full \gls{dd} domain representation~\cite{10042436,9362336}, which is expressed as
\begin{equation}
    \begin{aligned}
        \mathbf{y}_i^\mathrm{DD} & \triangleq\mathbf{H}_{c,i}^{\mathrm{DD,con}}\mathbf{s}_c+\sum_{i=1}^{I}\mathbf{H}_{p,i}^{\mathrm{DD,con}}\mathbf{s}_{p,i}+\mathbf{n}_i,
        \label{eq:y_dd_pure}
    \end{aligned}
\end{equation}
where $\mathbf{H}_{k,i}^{\mathrm{DD,con}}=(\mathbf{F}_N \otimes \mathbf{I}_M) \mathbf{H}_{k,i}^{\mathrm{TD}} (\mathbf{p}_k\otimes \tilde{\boldsymbol{\Psi}}_k)$, $k\in\{c,\allowbreak p\allowbreak\}$.
Such a formulation couples the precoder with the channel response, which complicates the subsequent precoder design.}
\par \textit{Remark 2:
    Compared to \gls{rsma}-\gls{ofdm} in \cite{10236464}, \gls{rsma}-\gls{otfs} introduces extra Doppler domain, making each symbol in $\mathbf{s}_c$ and $\mathbf{s}_{p,i}$ spread over the entire two-dimensional \gls{tf} plane. 
    This results in the multi-antenna precoding vector $\mathbf{p}_j$ being expanded to the Kronecker-structured precoder $\tilde{\mathbf{P}}_j$, which requires joint waveform shaping across antennas and across the whole two-dimensional \gls{otfs} frame.
    In effect, each data stream is embedded in the entire frame rather than confined to a particular subcarrier or time slot, which improves robustness under rapidly time-varying \gls{leo} satellite channels. }



\subsection{Channel Model}
The \gls{dd} domain channel between the satellite and user-$i$ is modeled as a sum of \gls{los} and \gls{nlos} paths \cite{9928043,9110855}, which is expressed as
\begin{equation}
    \begin{aligned}
        &\mathbf{h}_i^{DD}  (\tau,\nu)=\sqrt{\frac{\gamma_i}{\gamma_i+1}}h_i^\mathrm{LoS}\mathbf{a}^T(\theta_{i}^\mathrm{LoS})\delta\left(\tau-\tau_i^\mathrm{LoS}\right)\delta\left(\nu-\nu_i^\mathrm{LoS}\right)                        \\
                          &\, +\sqrt{\frac{1}{\gamma_i+1}}\sum_{q=1}^{Q_i}h_{i,q}^{\mathrm{NLoS}}\mathbf{a}^T(\theta_{i,q}^\mathrm{NLoS})\delta\left(\tau-\tau_{i,q}^{\mathrm{NLoS}}\right)\delta\left(\nu-\nu_{i,q}^{\mathrm{NLoS}}\right),
    \end{aligned}
\end{equation}
where the \gls{los} component is characterized by complex path gain~$ h_i^\mathrm{LoS} $, \gls{aod} $\theta_{i}^\mathrm{LoS}$, propagation delay~$ \tau_i^{\mathrm{LoS}} $, and Doppler shift~$ \nu_i^{\mathrm{LoS}} $; the~$ q $-th \gls{nlos} path is described by~$ h_{i,q}^{\mathrm{NLoS}} $, $\theta_{i,q}^\mathrm{NLoS}$,~$ \tau_{i,q}^{\mathrm{NLoS}} $, and~$ \nu_{i,q}^{\mathrm{NLoS}} $, representing the multipath gain, \gls{aod}, propagation delay, and Doppler shift, respectively;~$ \gamma_i $ denotes the Rician factor that governs the power ratio between the \gls{los} and scattered components;~$ Q_i $ is the total number of \gls{nlos} paths; and~$ \mathbf{a} \in \mathbb{C}^{N_t \times 1} $ denotes the spatial array steering vector of the \gls{ula} for the \gls{leo} satellite. The parameters are elaborated upon in the subsequent discussion.

\subsubsection{Path gain}
The terrestrial-satellite link is characterized by sparse scattering, leading to a dominant \gls{los} component and weak \gls{nlos} paths, which are captured by Rician factor~$ \gamma_i $. Denote the \gls{los} gain as~$ h_i^{\mathrm{LoS}} = e^{j\varphi_i} $ with~$ \varphi_i \sim \mathcal{U}(0,2\pi) $, and \gls{nlos} gains follow~$ h_{i,q}^{\mathrm{NLoS}} \sim \mathcal{CN}(0,\sigma_{i,q}^2) $.


\subsubsection{Spatial array steering vector}
The \gls{ula} steering vector is given by
\begin{equation}
    \mathbf{a}(\theta_{i,q}) = \left[ e^{-j2\pi \frac{d}{\lambda} \sin(\theta_{i,q})\mathbf{n}_t}  \right]^T,
    \label{eq:steering}
\end{equation}
where~$ \mathbf{n}_t = [0,1,\dots,N_t-1] $, $\lambda$ is the carrier wavelength, and $d=\lambda/2$ is the antenna spacing.

\subsubsection{Doppler shift and Delay}
The Doppler shift $ \nu_{i}^\mathrm{LoS} $, associated with the \gls{los} component of user-$ i $ is dominated by two independent Doppler shifts, $ \nu_{i}^{\mathrm{LoS-S}} $ and $ \nu_{i}^{\mathrm{LoS-U}} $, which result from the movements of the \gls{leo} satellite and user, respectively. Given the high velocity of the \gls{leo} satellite, the Doppler shift, $ \nu_i^{\mathrm{LoS}} $, is primarily determined by the satellite's motion, leading to $ \nu_i^{\mathrm{LoS}} \approx \nu_i^{\mathrm{LoS-S}} $. Similarly, we have the Doppler shifts of \gls{nlos} paths, $ \nu_{i,q}^{\mathrm{NLoS}} $, are largely governed by the satellite's velocity, resulting in $ \nu_{i,q}^{\mathrm{NLoS}} \approx \nu_{i,q}^{\mathrm{NLoS-S}} $. 
We assume that most of the propagation delay is compensated through configuring the timing advance information  at each user, and the remaining delay offset is limited to the symbol level~\cite{38811,10959036}.

\par For notational simplicity, the \gls{los} path is indexed as~$ q=0 $, with effective gain~$ h_{i,0} = \sqrt{\frac{\gamma_i}{\gamma_i+1}} h_i^{\mathrm{LoS}} $, $\theta_{i,0}=\theta_{i}^\mathrm{LoS}$, delay~$ \tau_{i,0} = \tau_i^{\mathrm{LoS}} $ and Doppler shift~$ \nu_{i,0} = \nu_i^{\mathrm{LoS}} $. For~$ q = 1,\dots,Q_i $, \gls{nlos} components are defined as~$ h_{i,q} = \sqrt{\frac{1}{\gamma_i+1}} h_{i,q}^{\mathrm{NLoS}} $, $\theta_{i,q}=\theta_{i,q}^\mathrm{NLoS}$, $ \tau_{i,q} = \tau_{i,q}^{\mathrm{NLoS}} $ and $ \nu_{i,q} = \nu_{i,q}^{\mathrm{NLoS}} $. The \gls{dd} domain channel is expressed as
\begin{equation}
    \mathbf{h}_i^{\mathrm{DD}}(\tau, \nu) = \sum_{q=0}^{Q_i} h_{i,q} \mathbf{a}(\theta_{i,q})\delta(\tau - \tau_{i,q}) \delta(\nu - \nu_{i,q}) .
    \label{eq:channel_dd}
\end{equation}
To better capture practical channel characteristics, both fractional delay, and fractional Doppler shifts are considered. Accordingly, the delay and Doppler shift of the $q$-th path for the $i$-th user are expressed as
\begin{equation}
    \tau_{i,q} = \frac{l_{i,q} + \ell_{i,q}}{M\Delta f}, \quad
    \nu_{i,q} = \frac{k_{i,q} + \kappa_{i,q}}{NT},
    \label{eq:delay_doppler_indices}
\end{equation}
where $l_{i,q}$ and $k_{i,q}$ denote the integer delay and Doppler indices, while
$\ell_{i,q}, \kappa_{i,q} \in (-\tfrac{1}{2}, \tfrac{1}{2})$ represent the corresponding fractional components.
We assume the maximal delay and maximal Doppler of the propagation path is bounded, i.e., 
$\tau_{\max} \leq T$ and $\nu_{\max} \leq \Delta f$.
\par The \gls{td} domain channel corresponding to~\eqref{eq:channel_dd} can be expressed in an equivalent matrix form as 
\begin{equation}
    \mathbf{H}^{\mathrm{TD}}_i =  \sum_{q=0}^{Q_i} h_{i,q}e^{-j2\pi\frac{(k_{i,q} + \kappa_{i,q})(l_{i,q} + \ell_{i,q})}{MN}}\left[\mathbf{a}(\theta_{i,q})\otimes\breve{\mathbf{H}}^{\mathrm{TD}}_{i,q}\right],
    \label{eq:h_td_matrix}
\end{equation}
where $\breve{\mathbf{H}}_{i,q}^\mathrm{TD} = \boldsymbol{\Delta}_{i,q} \boldsymbol{\Xi}_{i,q}$, and the matrices $ \boldsymbol{\Delta}_{i,q} \in \mathbb{C}^{MN \times MN} $ and $ \boldsymbol{\Xi}_{i,q} \in \mathbb{C}^{MN \times MN} $ capture the effects of Doppler shifts and delay, respectively \cite{10283795}. The Doppler effects matrix $ \boldsymbol{\Delta}_{i,q} $ is a diagonal matrix with diagonal elements defined as
\begin{equation}
    \begin{aligned}
        &[\boldsymbol{\Delta}_{i,q}]_{n' + Nm', n' + Nm'}                                                                  \\
                                  &\quad = \exp{\left( j2\pi \frac{k_{i,q} + \kappa_{i,q}}{N}  (n' + \frac{m'}{M}) \right)}.
    \end{aligned}
    \label{eq:delta_diag}
\end{equation}
The $(n' + Nm', n + Nm)$-th element of the delay effects matrix $ \boldsymbol{\Xi}_{i,q} $ is
\begin{equation}
    \begin{aligned}
         &[\boldsymbol{\Xi}_{i,q}]_{n' + Nm', n + Nm}                                                      \\
                               &\quad = \mathrm{sinc}\left(M(n' - n) + m' - m - l_{q,i} - \ell_{q,i}\right),
    \end{aligned}
    \label{eq:gamma_sinc}
\end{equation}
The overall channel response across all users is represented as
\begin{equation}
    \mathbf{H}^\mathrm{TD} = [\mathbf{H}_1^\mathrm{TD}, \mathbf{H}_2^\mathrm{TD}, \dots, \mathbf{H}_I^\mathrm{TD}],
\end{equation}
where $\mathbf{H}_i^\mathrm{TD}$ denotes the channel matrix associated with the $i$-th user. 

\subsection{Channel State Information}
In practical satellite communication systems, perfect \gls{csi} acquisition is infeasible due to several factors such as hardware impairments, and estimation errors~\cite{7892949,8671740}. Consequently, the satellite (transmitter) has access only to a partial and imperfect channel knowledge.
The actual channel is modeled as the sum of the estimated channel and an error component\footnote{For notational simplicity, the domain superscripts (e.g., \gls{dd} or \gls{td}) of the channel are omitted throughout this subsection.} \cite{7390017,10493204}
\begin{equation}
    \mathbf{H}_i = \hat{\mathbf{H}}_i + {\mathbf{E}}_{h,i},
\end{equation}
where $\mathbf{E}_{h,i}$ captures the estimation uncertainty for user~$i$, modeled as
\begin{equation}
    \mathbf{E}_{h,i}=\sum_{q=0}^{Q_i} e_{i,q}e^{-j2\pi\frac{(k_{i,q} + \kappa_{i,q})(l_{i,q} + \ell_{i,q})}{MN}}
    \left[\mathbf{a}(\theta_{i,q})\otimes\breve{\mathbf{H}}_{i}\right]
\end{equation}
with $e_{i,q}\sim \mathcal{CN}(0, \sigma_{h,ei}^2)$ denoting the estimation error of the $q$-th path gain for user~$i$. The variance of the path-gain estimation error is given by $\sigma_{h,ei}^2=\rho^2\gamma_e+\frac{1-\rho^2}{Q_i}$, where $\rho$ is the correlation coefficient between the estimated and actual channel gains, and $\gamma_e$ denotes the \gls{snr} of the pilot signal.

\par We further assume that the joint distribution $(\mathbf{H}, \hat{\mathbf{H}})$ is stationary and ergodic~\cite{7555358}, and that the conditional distribution $f_{\mathbf{H}|\hat{\mathbf{H}}}(\mathbf{H} \mid \hat{\mathbf{H}})$ is known at the transmitter side, while the instantaneous realization of $\hat{\mathbf{H}}$ remains unknown during transmission, corresponding to statistical \gls{csit} and perfect \gls{csir} conditions. In this work, we mainly focus on the estimation errors that occur in the path gains, $e_{i,q}$.

\subsection{Achievable Rate}
\par At user $i$, signal detection is carried out following the \gls{sic} strategy. Specifically, the user first decodes the common stream vector $\mathbf{s}_c$ while treating interference from all private streams as noise. Once the common stream is successfully recovered and removed, user~$i$ proceeds to decode its intended private stream vector $\mathbf{s}_{p,i}$, where the remaining interference from other users’ private streams is regarded as noise. The recovered message for user~$i$ is obtained by combining the decoded portion of the common stream with its private stream. In the following, we derive the instantaneous and achievable ergodic rate for both the common and private streams, referred as the common rate and private rate, respectively.

\par The covariance matrices of the received signals corresponding to the common and private streams of user~$i$ are defined in \eqref{eq:T_c_i} and \eqref{eq:R_p_i}, respectively, as shown at the top of the this page.
\begin{figure*}[t!]
    \begin{subequations}
        \begin{align}
            \mathbf{T}_{c,i}
             & \triangleq \mathbb{E}\left\{\mathbf{y}_i^\mathrm{DD}{\mathbf{y}_i^\mathrm{DD}}^H\right\}=
            \underbrace{{\tilde{\mathbf{H}}_i^{\mathrm{DD}}} \tilde{\mathbf{P}}_c \tilde{\mathbf{P}}_c^H {(\tilde{\mathbf{H}}_i^{\mathrm{DD}})}^H}_{\mathbf{S}_{c,i}} + \underbrace{\sum_{i^{\prime}=1}^{I} \tilde{\mathbf{H}}_{i}^\mathrm{DD} \tilde{\mathbf{P}}_{p,i^{\prime}} {\tilde{\mathbf{P}}_{p,i^{\prime}}}^H {(\tilde{\mathbf{H}}_{i}^\mathrm{DD})}^H + \mathbf{I}_{MN}}_{\mathbf{K}_{c,i}}, \label{eq:T_c_i} \\
            \mathbf{T}_{p,i}
             & \triangleq \mathbb{E}\left\{(\mathbf{y}_i^\mathrm{DD} - \tilde{\mathbf{H}}_i^\mathrm{DD} \tilde{\mathbf{P}}_c {\mathbf{s}}_c)(\mathbf{y}_i^\mathrm{DD} - \tilde{\mathbf{H}}_i^\mathrm{DD} \tilde{\mathbf{P}}_c {\mathbf{s}}_c)^H\right\}=
            \underbrace{{\tilde{\mathbf{H}}_i^{\mathrm{DD}}} \tilde{\mathbf{P}}_{p,i} \tilde{\mathbf{P}}_{p,i}^H {(\tilde{\mathbf{H}}_i^{\mathrm{DD}})}^H}_{\mathbf{S}_{p,i}} + \underbrace{\sum_{i^{\prime}=1,\,i^{\prime}\neq i}^{I} \tilde{\mathbf{H}}_{i}^\mathrm{DD} \tilde{\mathbf{P}}_{p,i^{\prime}} {\tilde{\mathbf{P}}_{p,i^{\prime}}}^H {(\tilde{\mathbf{H}}_{i}^\mathrm{DD})}^H + \mathbf{I}_{MN}}_{\mathbf{K}_{p,i}}. \label{eq:R_p_i}
        \end{align}
    \end{subequations}
    \hrulefill
\end{figure*}

\par Since the transmitter determines its precoders based on the estimated channel state $\hat{\mathbf{H}}_i^\mathrm{DD}$, while each user decodes its signals using the actual channel realization $\mathbf{H}_i^\mathrm{DD}$ with perfect \gls{csir}, the pair $(\hat{\mathbf{H}}_i^\mathrm{DD}, \mathbf{H}_i^\mathrm{DD})$ jointly determines the instantaneous achievable rates of the common and private streams. Assuming Gaussian signaling, the instantaneous \gls{sinr} of the common and private streams at user~$i$ are given by
\begin{subequations}
    \begin{align}
        \boldsymbol{\Gamma}_{c,i} & = \tilde{\mathbf{P}}_c^H ({\tilde{\mathbf{H}}_i^\mathrm{DD})}^H \mathbf{K}_{c,i}^{-1} \tilde{\mathbf{H}}_i^\mathrm{DD} \tilde{\mathbf{P}}_c,           \\
        \boldsymbol{\Gamma}_{p,i} & = \tilde{\mathbf{P}}_{p,i}^H {(\tilde{\mathbf{H}}_i^\mathrm{DD})}^H \mathbf{K}_{p,i}^{-1} {\tilde{\mathbf{H}}_i^\mathrm{DD}} \tilde{\mathbf{P}}_{p,i}.
    \end{align}
\end{subequations}
the corresponding instantaneous achievable rates for user~$i$ are expressed as
\begin{subequations}
    \begin{align}
        R_{c,i} & = \frac{1}{MN}\log_2 \det\left( \mathbf{I}_{MN} + \boldsymbol{\Gamma}_{c,i} \right), \label{eq:R_c_k} \\
        R_{p,i} & = \frac{1}{MN}\log_2 \det\left( \mathbf{I}_{MN} + \boldsymbol{\Gamma}_{p,i} \right). \label{eq:R_p_k}
    \end{align}
\end{subequations}
Since the transmitter only has partial \gls{csit}, relying on instantaneous rates may overestimate system performance~\cite{7555358}. To capture the average performance, we define ergodic rate of the common and private streams as
\begin{equation}
    \bar{R}_{c,i} = \mathbb{E}_{(\mathbf{H}_i^\mathrm{DD}, \hat{\mathbf{H}}_i^\mathrm{DD})} \left\{ R_{c,i} \right\}, \,
    \bar{R}_{p,i} = \mathbb{E}_{(\mathbf{H}_i^\mathrm{DD}, \hat{\mathbf{H}}_i^\mathrm{DD})} \left\{ R_{p,i} \right\},
\end{equation}
where the expectation is taken over the joint distribution of $(\mathbf{H}_i^\mathrm{DD}, \hat{\mathbf{H}}_i^\mathrm{DD})$. To ensure that all users can decode the common stream, the common ergodic rate must satisfy $\bar{R}_c = \min_{i \in \mathcal{I}} \bar{R}_{c,i}$. The common rate $\bar{R}_c$ is shared among users as $\sum_{i \in \mathcal{I}} \bar{C}_i = \bar{R}_c$, and the total ergodic rate achieved by user~$i$ is given by
$\bar{R}_{\mathrm{tot},i} = \bar{C}_i + \bar{R}_{p,i}.$

\section{Problem formulation}\label{sec:problem_formulation}

In this section, we formulate an \gls{esr} maximization problem for the considered system.
\par To illustrate the impact of channel uncertainty, we first consider a baseline strategy where the transmitter relies on the estimated channel $\hat{\mathbf{H}}_i^{\mathrm{DD}}$ and determines the instantaneous precoder $\mathbf{P}_{\mathrm{BF}}$ by maximizing the instantaneous sum-rate under the power constraint $\mathrm{tr}(\mathbf{P}_{\mathrm{BF}}\mathbf{P}_{\mathrm{BF}}^{H}) \leq P_\mathrm{t}$.
While this approach is intuitive, it implicitly assumes that the channel estimate is accurate. In practice, however, estimation errors create a mismatch between the adopted precoder and the actual channel realization. Such mismatch is particularly detrimental in multi-user settings, where inaccurate channel knowledge may cause the transmitter to select rates that cannot be reliably decoded, resulting in performance degradation~\cite{7555358}.
These considerations motivate an optimization framework that explicitly accounts for \gls{csit} uncertainty when determining the transmission strategy.

\par To capture the long-term behavior of the system under such uncertainty, we focus on maximizing the \gls{esr}, which reflects the achievable performance averaged over all channel realizations and provides a more robust performance measure than instantaneous rate optimization. For the \gls{rsma}-\gls{otfs} transmission strategy, the \gls{esr} is denoted as $\bar{R} = \sum_{i=1}^{I} \bar{R}_{\mathrm{tot},i}$. 
Since the instantaneous rate cannot be perfectly obtained under imperfect \gls{csit}, the transmitter relies on the average rate, which represents the short-term expectation conditioned on the estimated channel.

For a given channel estimate $ \hat{\mathbf{H}}_i^{\mathrm{DD}} $ and its corresponding precoder $ \mathbf{P}_\mathrm{BF}(\hat{\mathbf{H}}_i^{\mathrm{DD}}) $, the conditional average rate is defined as the conditional expectation of the instantaneous rate with respect to the \gls{csit} error distribution:
\begin{subequations}
    \begin{align}
        \hat{R}_{c,i}(\hat{\mathbf{H}}_i^\mathrm{DD}) & =
        \mathbb{E}_{\mathbf{H}_i^\mathrm{DD}|\hat{\mathbf{H}}_i^\mathrm{DD}}
        \left\{ R_{c,i}(\mathbf{H}_i^\mathrm{DD}, \hat{\mathbf{H}}_i^\mathrm{DD}) \mid \hat{\mathbf{H}}_i^\mathrm{DD} \right\}, \\
        \hat{R}_{p,i}(\hat{\mathbf{H}}_i^\mathrm{DD}) & =
        \mathbb{E}_{\mathbf{H}_i^\mathrm{DD}|\hat{\mathbf{H}}_i^\mathrm{DD}}
        \left\{ R_{p,i}(\mathbf{H}_i^\mathrm{DD}, \hat{\mathbf{H}}_i^\mathrm{DD}) \mid \hat{\mathbf{H}}_i^\mathrm{DD} \right\}.
    \end{align}
\end{subequations}

\par The conditional average rate differs from the ergodic rate because the former represents the expected rate under a specific channel estimate, while the latter characterizes the long-term average over all joint channel realizations. According to the law of total expectation, their relationship can be expressed as        $\bar{R}_{c,i}  =
    \mathbb{E}_{\hat{\mathbf{H}}_i^\mathrm{DD}}
    \left\{ \hat{R}_{c,i}(\hat{\mathbf{H}}_i^\mathrm{DD}) \right\}$,
$\bar{R}_{p,i}  =
    \mathbb{E}_{\hat{\mathbf{H}}_i^\mathrm{DD}}
    \left\{ \hat{R}_{p,i}(\hat{\mathbf{H}}_i^\mathrm{DD}) \right\}$.
For the common stream, let $ \hat{C}_i $ denote the average rate portion of user-$ i $, satisfying $\sum_{i=1}^{I} \hat{C}_i = \hat{R}_c$, where $ \hat{R}_c \leq \min_{i \in \mathcal{I}} \{ \hat{R}_{c,i} \} $. Moreover, the inequality
$\min_{i \in \mathcal{I}}
    \left\{ \mathbb{E}_{\hat{\mathbf{H}}_i^{\mathrm{DD}}} [ \hat{R}_{c,i} ] \right\}
    \geq
    \mathbb{E}_{\hat{\mathbf{H}}_i^{\mathrm{DD}}}
    \left[ \min_{i \in \mathcal{I}} \{ \hat{R}_{c,i} \} \right]$,
follows from the fact that exchanging the expectation and minimization operations cannot increase the result \cite{7555358}.

By invoking the \gls{lln}, the ergodic rate can be approximated by averaging the average rate over all estimated channel states, thus decoupling dependencies across channel realizations. Consequently, the long-term \gls{esr} maximization problem under imperfect \gls{csit} can be reformulated as a short-term \gls{asr} optimization for each channel estimate $ \hat{\mathbf{H}}_i^{\mathrm{DD}} $, expressed as
\begin{subequations}
    \label{prob:asr}
    \begin{align}
        \mathcal{P}_{1}:\,\max_{\mathbf{P}_\mathrm{BF}, \boldsymbol{\Psi}, \hat{\mathbf{c}}} & \, \sum_{i=1}^{I} \hat{R}_{\mathrm{tot},i},                                                                                                                                  \\
        \text{s.t.} \,\,                                                                     & \,\sum_{i^\prime=1}^{I} \hat{C}_{i^\prime} \leq \hat{R}_{c,i}^{(L)},\,i \in \mathcal{I},                                                                                     \\
                                                                                             & \, \tr\left(\mathbf{P}_\mathrm{BF}\mathbf{P}_\mathrm{BF}^H\right)  \leq P_\mathrm{t},\label{prob:asr_c}                                                                      \\
                                                                                             & \, \boldsymbol{\Psi}_\star \in \boldsymbol{\Psi},\,[\boldsymbol{\Psi}_\star]_{ij} = 0 \text{ for } i \ne j, \, [\boldsymbol{\Psi}_\star]_{ii} \in \{0,1\},\label{prob:asr_d} \\
                                                                                             & \,\hat{\mathbf{c}} \geq \mathbf{0},\label{prob:asr_e}
    \end{align}
\end{subequations}
where $ \hat{\mathbf{c}} = [\hat{C}_1, \hat{C}_2, \dots, \hat{C}_I] $ denotes the vector of average common-rate allocations, and $ \hat{R}_{\mathrm{tot},i} = \hat{C}_i  + \hat{R}_{p,i}$ is the total average rate achieved by user~$i$.

\section{Proposed Algorithm} \label{sec:proposed_algorithm}
The optimization problem in \eqref{prob:asr} is non-deterministic, making a direct solution intractable. To address this issue, we propose a three-stage procedure. First, the \gls{saa} method is used to obtain a deterministic approximation of the original statistical optimization problem. Second, the resultant \gls{asr} problem is reformulated as an average \gls{awmse} problem to efficiently address the rate maximization objective. Finally, vectorization transforms the \gls{awmse} problem into a tractable formulation to avoid matrix inversion, whose sequential problem could finally be solved using \gls{ao}. 

\subsection{Sample Average Approximation}
Consider a set of $ L $ \gls{iid} channel realizations indexed by $ \mathcal{L} = \{1, 2, \dots, L\} $, drawn from the conditional distribution with density $ f_{\mathbf{H}_i^\mathrm{DD}|\hat{\mathbf{H}}_i^\mathrm{DD}}\left( \mathbf{H}_i^\mathrm{DD} \mid \hat{\mathbf{H}}_i^\mathrm{DD} \right) $. For a specific estimated channel $ \hat{\mathbf{H}}_i^\mathrm{DD} $, the sampled channel is given by
\begin{equation}
    {\mathbb{H}_i^\mathrm{DD}}^{(L)} \triangleq
    \left\{ {\mathbf{H}_i^\mathrm{DD}}^{(l)} = \hat{\mathbf{H}}_i^\mathrm{DD} + \mathbf{E}_{h,i}^{(l)}
    \mid \hat{\mathbf{H}}_i^\mathrm{DD},\, l \in \mathcal{L} \right\}.
\end{equation}
Based on the channel realizations, we define the \gls{saf} of the average rates as
$\hat{R}_{c,i}^{(L)}  \triangleq
    \frac{1}{L} \sum_{l=1}^{L} R_{c,i}^{(l)}, \,
    \hat{R}_{p,i}^{(L)}  \triangleq
    \frac{1}{L} \sum_{l=1}^{L} R_{p,i}^{(l)},$
where $ R_{z,i}^{(l)} \triangleq R_{z,i}({\mathbf{H}_i^\mathrm{DD}}^{(l)}) $, $z\in\{c,p\}$ represents the instantaneous common or private rate corresponding to the $l$-th  channel realization.

The \gls{saa} formulation of the stochastic optimization problem~\eqref{prob:asr} can thus be written as
\begin{subequations}
    \label{prob:saa}
    \begin{align}
        \mathcal{P}_{2}:\, \max_{\mathbf{P}_\mathrm{BF}, \boldsymbol{\Psi}, \hat{\mathbf{c}}}
         & \quad \sum_{i=1}^{I} \hat{R}_{\mathrm{tot},i}^{(L)}                                                \\
        \text{s.t.} \,\,
         & \quad\sum_{i^\prime=1}^{I} \hat{C}_{i^\prime}        \leq \hat{R}_{c,i}^{(L)},\,i \in \mathcal{I}, \\
         & \quad \eqref{prob:asr_c},\,\eqref{prob:asr_d},\,\eqref{prob:asr_e},
    \end{align}
\end{subequations}
where $ \hat{R}_{\mathrm{tot},i}^{(L)} = \hat{C}_i + \hat{R}_{p,i}^{(L)} $. Since the achievable rates are bounded~\cite{7555358}, the \gls{lln} ensures that
\begin{subequations}
    \label{eq:SAA_LLN_rate1}
    \begin{align}
        \lim_{L \to \infty}
        \hat{R}_{c,i}^{(L)}(\mathbf{P}_\mathrm{BF},\boldsymbol{\Psi})
         & = \hat{R}_{c,i}(\mathbf{P}_\mathrm{BF},\boldsymbol{\Psi}),
        \, \forall\, \mathbf{P}_\mathrm{BF} \in \mathbb{P}_\mathrm{BF},\boldsymbol{\Psi}\in\mathbb{A}_\mathrm{M} \\
        \lim_{L \to \infty}
        \hat{R}_{p,i}^{(L)}(\mathbf{P}_\mathrm{BF},\boldsymbol{\Psi})
         & = \hat{R}_{p,i}(\mathbf{P}_\mathrm{BF},\boldsymbol{\Psi}),
        \, \forall\, \mathbf{P}_\mathrm{BF} \in \mathbb{P}_\mathrm{BF},\boldsymbol{\Psi}\in\mathbb{A}_\mathrm{M},
    \end{align}
\end{subequations}
where $ \mathbb{P}_\mathrm{BF} \triangleq
    \left\{ \mathbf{P}_\mathrm{BF} \in \mathbb{C}^{(I+1)\times N_t} \mid
    \tr(\mathbf{P}_\mathrm{BF}\mathbf{P}_\mathrm{BF}^H) \leq P_\mathrm{t} \right\} $ and $\mathbb{A}_\mathrm{M}\triangleq\allowbreak\left\{\boldsymbol{\Psi}_\star|\allowbreak\boldsymbol{\Psi}_\star\in\boldsymbol{\Psi},\,[\boldsymbol{\Psi}_\star]_{ii} \allowbreak\in \{0,1\}, \,[\boldsymbol{\Psi}_\star]_{ij}=0 \,\mathrm{for}\, i\ne j\right\}$ denote the feasible set of precoders and arrangement matrices. Within these sets, the rate functions are bounded, continuous, and differentiable with respect to $ \mathbf{P}_\mathrm{BF} $, ensuring the convergence in~\eqref{eq:SAA_LLN_rate1}. Therefore, as $ L \to \infty $, $\hat{R}_{z,i}^{(L)}$ and $\hat{R}_{z,i}$, $ z \in \{c,p\} $, become asymptotically equivalent, leading to
\begin{equation}
    \lim_{L \to \infty}
    \hat{R}_{\mathrm{tot},i}^{(L)}
    = \hat{R}_{\mathrm{tot},i},
    \quad \forall\, \mathbf{P}_\mathrm{BF} \in \mathbb{P}_\mathrm{BF},\,\boldsymbol{\Psi}\in\mathbb{A}_\mathrm{M}.
    \label{eq:SAA_LLN_rate2}
\end{equation}
From \eqref{eq:SAA_LLN_rate1} and~\eqref{eq:SAA_LLN_rate2}, we obtain that the global optimal solution of the \gls{saa} problem in~\eqref{prob:saa} converges to that of the original stochastic optimization problem in~\eqref{prob:asr} as $ L \to \infty $~\cite{7555358, Shapiro2021}. We then proceed to solve the sampled problem in \eqref{prob:saa}. 

\subsection{Average Augmented Weighted MSE Minimization}
To efficiently solve the \gls{saa} problem in~\eqref{prob:saa}, we reformulate it into an equivalent \gls{awmse} representation for tractability, following~\cite{4712693}. \gls{awmse} formulations differ from conventional \gls{wmse} counterparts, since the weighted factors are treated as optimization variables rather than fixed parameters, and the objective function is modified to include the logarithmic terms of the weights. Then, the relationship between the achievable rate and the \gls{wmmse} is established based on the aforementioned features.

\par Let $\mathbf{A}_{c,i}, \mathbf{A}_{p,i} \in \mathbb{C}^{MN \times MN}$ denote the receive equalizers (filters) for the common and private streams of user~$i$, respectively. The estimated common stream is $\hat{\mathbf{s}}_{c,i} = \mathbf{A}_{c,i}\mathbf{y}_i^\mathrm{DD}$. Once the common stream $\mathbf{s}_c$ is successfully decoded and subtracted, the private stream of user~$i$ is estimated as $\hat{\mathbf{s}}_{p,i} = \mathbf{A}_{p,i} (\mathbf{y}_i^\mathrm{DD} - \mathbf{H}_i^\mathrm{DD} \tilde{\mathbf{P}}_c {\mathbf{s}}_c)$.

\par The \gls{mse} matrices of the common and private streams are expressed in \eqref{eq:E_c_i} and \eqref{eq:E_p_i} at the top of next page.
\begin{figure*}[t!]
    \begin{subequations}
        \begin{align}
            \mathbf{E}_{c,i} & = \mathbb{E} \left\{ (\hat{\mathbf{s}}_{c,i} - \mathbf{s}_c)(\hat{\mathbf{s}}_{c,i} - \mathbf{s}_c)^H \right\} =\mathbf{A}_{c,i} \mathbf{y}_i^\mathrm{DD}{\mathbf{y}_i^\mathrm{DD}}^H \mathbf{A}_{c,i}^H-\mathbf{A}_{c,i}\mathbf{y}_i^\mathrm{DD}\mathbf{s}_c^H-\mathbf{s}_c {\mathbf{y}_i^\mathrm{DD}}^H \mathbf{A}_{c,i}^H + \mathbf{I}_{MN} \label{eq:E_c_i}                \\
            \mathbf{E}_{p,i} & = \mathbb{E} \left\{ (\hat{\mathbf{s}}_{p,i} - \mathbf{s}_{p,i})(\hat{\mathbf{s}}_{p,i} - \mathbf{s}_{p,i})^H \right\}=\mathbf{A}_{p,i} \mathbf{y}_i^\mathrm{DD}{\mathbf{y}_i^\mathrm{DD}}^H \mathbf{A}_{p,i}^H-\mathbf{A}_{p,i}\mathbf{y}_i^\mathrm{DD}\mathbf{s}_{p,i}^H-\mathbf{s}_{p,i} {\mathbf{y}_i^\mathrm{DD}}^H \mathbf{A}_{p,i}^H + \mathbf{I}_{MN} \label{eq:E_p_i}
        \end{align}
    \end{subequations}
    \hrulefill
\end{figure*}
The optimum \gls{mmse} equalizers are derived by minimizing the corresponding \gls{mse} expressions with respect to the receive filters, i.e.,
$\frac{\partial \mathbf{E}_{c,i}}{\partial \mathbf{A}_{c,i}} = 0, \, \frac{\partial \mathbf{E}_{p,i}}{\partial \mathbf{A}_{p,i}} = 0,$
which lead to
\begin{subequations}
    \begin{align}
        \mathbf{A}_{c,i}^{\mathrm{MMSE}} & = \arg\min_{\mathbf{A}_{c,i}}\mathbb{E}\left\{||\mathbf{A}_{c,i}\mathbf{y}_i^\mathrm{DD}-\mathbf{s}_c||^2\right\}\notag                                                                                          \\
                                         & =\tilde{\mathbf{P}}_c^H {(\tilde{\mathbf{H}}_i^\mathrm{DD})}^H \mathbf{T}_{c,i}^{-1},          \label{eq:mmse_c}                                                                                                 \\
        \mathbf{A}_{p,i}^{\mathrm{MMSE}} & =\arg\min_{\mathbf{A}_{p,i}}\mathbb{E}\left\{||\mathbf{A}_{p,i} \left(\mathbf{y}_i^\mathrm{DD} - \tilde{\mathbf{H}}_i^\mathrm{DD} \tilde{\mathbf{P}}_c {\mathbf{s}}_c\right)-\mathbf{s}_{p,i}||^2\right\} \notag \\
                                         & =\tilde{ \mathbf{P}}_{p,i}^H {(\tilde{\mathbf{H}}_i^\mathrm{DD})}^H \mathbf{T}_{p,i}^{-1}.\label{eq:mmse_p}
    \end{align}
\end{subequations}

\par Substituting \eqref{eq:mmse_c} and \eqref{eq:mmse_p} into \eqref{eq:E_c_i} and \eqref{eq:E_p_i}, respectively, the corresponding \gls{mmse} covariance matrices are obtained as
\begin{subequations}
    \label{eq:opt_mmse}
    \begin{align}
        \mathbf{E}_{c,i}^{\mathrm{MMSE}} & \triangleq \min_{\mathbf{A}_{c,i}} \mathbf{E}_{c,i} = \left( \mathbf{I}_{MN} + \tilde{\mathbf{P}}_c^H {(\tilde{\mathbf{H}}_i^\mathrm{DD})}^H \mathbf{K}_{c,i}^{-1} \tilde{\mathbf{H}}_i^\mathrm{DD} \tilde{\mathbf{P}}_c \right)^{-1},          \\
        \mathbf{E}_{p,i}^{\mathrm{MMSE}} & \triangleq \min_{\mathbf{A}_{p,i}} \mathbf{E}_{p,i}  = \left( \mathbf{I}_{MN} + \tilde{\mathbf{P}}_{p,i}^H {(\tilde{\mathbf{H}}_i^\mathrm{DD})}^H \mathbf{K}_{p,i}^{-1} \tilde{\mathbf{H}}_i^\mathrm{DD} \tilde{\mathbf{P}}_{p,i} \right)^{-1}.
    \end{align}
\end{subequations}
Accordingly, the instantaneous achievable rates are obtained as
\begin{equation}
    R_{c,i}  = \log_2 \det   \left(\mathbf{E}_{c,i}^{\mathrm{MMSE}}\right)^{-1} , \,
    R_{p,i}  = \log_2 \det  \left(\mathbf{E}_{p,i}^{\mathrm{MMSE}}\right)^{-1} .
\end{equation}
By introducing weight matrices $\mathbf{B}_{c,i}$ and $\mathbf{B}_{p,i}$, the \gls{awmse} for decoding common and private streams at user-$i$ are expressed as
\begin{subequations}
    \label{eq:awmse}
    \begin{align}
        {\xi}_{c,i} & = \tr\left( \mathbf{B}_{c,i} \mathbf{E}_{c,i} \right) - \log \det (\mathbf{B}_{c,i}),\label{eq:awmse_common} \\
        {\xi}_{p,i} & = \tr\left( \mathbf{B}_{p,i} \mathbf{E}_{p,i} \right) - \log \det (\mathbf{B}_{p,i}).
    \end{align}
\end{subequations}
To establish the rate\textendash\gls{wmmse} relationship, we treat the equalizers and weights as optimization variables. Taking the derivative of ${\xi}_{c,i}$ and ${\xi}_{p,i}$ with respect to $\mathbf{A}_{c,i}$ and $\mathbf{A}_{p,i}$ and setting the gradients to zero yields $\frac{\partial {\xi}_{c,i}}{\partial \mathbf{A}_{c,i}} = 0,\,
    \frac{\partial {\xi}_{p,i}}{\partial \mathbf{A}_{p,i}} = 0$, which leads to the optimal equalizers
\begin{equation}
    \mathbf{A}_{c,i}^* = \mathbf{A}_{c,i}^{\mathrm{MMSE}}, \quad
    \mathbf{A}_{p,i}^* = \mathbf{A}_{p,i}^{\mathrm{MMSE}}.
\end{equation}
By substituting the \gls{mmse} equalizers into \eqref{eq:awmse}, the \gls{awmse} can be rewritten as
\begin{subequations}
    \label{eq:awmse_opt_mse}
    \begin{align}
        {\xi}_{c,i}(\mathbf{A}_{c,i}^{\mathrm{MMSE}}) & =
        \tr\left( \mathbf{B}_{c,i} \mathbf{E}_{c,i}^{\mathrm{MMSE}} \right)
        - \log \det (\mathbf{B}_{c,i}),                   \\
        {\xi}_{p,i}(\mathbf{A}_{p,i}^{\mathrm{MMSE}}) & =
        \tr\left( \mathbf{B}_{p,i} \mathbf{E}_{p,i}^{\mathrm{MMSE}} \right)
        - \log \det (\mathbf{B}_{p,i}).
    \end{align}
\end{subequations}
Next, by differentiating ${\xi}_{c,i}(\mathbf{A}_{c,i}^{\mathrm{MMSE}})$ and ${\xi}_{p,i}(\mathbf{A}_{p,i}^{\mathrm{MMSE}})$ with respect to the corresponding weight matrix and setting the gradient to zero, i.e.,  $\frac{\partial {\xi}_{c,i}(\mathbf{A}_{c,i}^{\mathrm{MMSE}})}{\partial \mathbf{B}_{c,i}} = 0,\quad
    \frac{\partial {\xi}_{p,i}(\mathbf{A}_{p,i}^{\mathrm{MMSE}})}{\partial \mathbf{B}_{p,i}} = 0, $ the optimal \gls{mmse} weights are obtained as
\begin{subequations}
    \label{eq:opt_weight}
    \begin{align}
        \mathbf{B}_{c,i}^* & = \mathbf{B}_{c,i}^{\mathrm{MMSE}} \triangleq (\mathbf{E}_{c,i}^{\mathrm{MMSE}})^{-1}, \\
        \mathbf{B}_{p,i}^* & = \mathbf{B}_{p,i}^{\mathrm{MMSE}} \triangleq (\mathbf{E}_{p,i}^{\mathrm{MMSE}})^{-1}.
    \end{align}
\end{subequations}
Finally, substituting the optimal weights into \eqref{eq:awmse_opt_mse} establishes the instantaneous rate\textendash\gls{wmmse} relationships
\begin{equation}
    \label{eq:rate-wmmse}
    \begin{aligned}
        {\xi}_{c,i} & \triangleq \min_{\mathbf{A}_{c,i}, \mathbf{B}_{c,i}} {\xi}_{c,i}
        = MN - R_{c,i},                                                                \\
        {\xi}_{p,i} & \triangleq \min_{\mathbf{A}_{p,i}, \mathbf{B}_{p,i}} {\xi}_{p,i}
        = MN - R_{p,i}.
    \end{aligned}
\end{equation}

\par Following the \gls{saa} framework, the average rate\textendash\gls{wmmse} relation is obtained by taking the expectation with respect to the conditional distribution of $\mathbf{H}_i^\mathrm{DD}$ given $\hat{\mathbf{H}}_i^\mathrm{DD}$, leading to
\begin{subequations}
    \label{eq:ar-wmmse_relationship}
    \begin{align}
        \hat{{\xi}}_{c,i} & \triangleq
        \mathbb{E}_{\mathbf{H}_i^\mathrm{DD}|\hat{\mathbf{H}}_i^\mathrm{DD}}
        \left\{ \min_{\mathbf{A}_{c,i}, \mathbf{B}_{c,i}}
        {\xi}_{c,i} \mid \hat{\mathbf{H}}_i^\mathrm{DD} \right\}
        = MN - \hat{R}_{c,i},          \\
        \hat{{\xi}}_{p,i} & \triangleq
        \mathbb{E}_{\mathbf{H}_i^\mathrm{DD}|\hat{\mathbf{H}}_i^\mathrm{DD}}
        \left\{ \min_{\mathbf{A}_{p,i}, \mathbf{B}_{p,i}}
        {\xi}_{p,i} \mid \hat{\mathbf{H}}_i^\mathrm{DD} \right\}= MN - \hat{R}_{p,i}.
    \end{align}
\end{subequations}
To obtain deterministic counterparts of~\eqref{eq:ar-wmmse_relationship}, the average \gls{awmse}s are approximated using the \gls{saf} method. Specifically,
\begin{equation}
    \hat{{\xi}}_{c,i}^{(L)} \triangleq
    \frac{1}{L} \sum_{l=1}^{L} {\xi}_{c,i}^{(l)},\quad
    \hat{{\xi}}_{p,i}^{(L)} \triangleq
    \frac{1}{L} \sum_{l=1}^{L} {\xi}_{p,i}^{(l)},
\end{equation}
where ${\xi}_{c,i}^{(l)}\triangleq {\xi_{c,i}}({\mathbf{H}_i^\mathrm{DD}}^{(l)},\mathbf{A}_{c,i}^{(l)},\mathbf{B}_{c,i}^{(l)})$,
${\xi}_{p,i}^{(l)}\triangleq {\xi_{p,i}}({\mathbf{H}_i^\mathrm{DD}}^{(l)},\allowbreak\mathbf{A}_{p,i}^{(l)},\mathbf{B}_{p,i}^{(l)})$,
$\mathbf{A}_{c,i}^{(l)}\triangleq \mathbf{A}_{c,i}^{(l)}({\mathbf{H}_i^\mathrm{DD}}^{(l)})$,
$\mathbf{A}_{p,i}^{(l)}\triangleq \mathbf{A}_{p,i}^{(l)}({\mathbf{H}_i^\mathrm{DD}}^{(l)})$,
$\mathbf{B}_{c,i}^{(l)}\triangleq \mathbf{B}_{c,i}^{(l)}({\mathbf{H}_i^\mathrm{DD}}^{(l)})$, and $\mathbf{B}_{p,i}^{(l)}\triangleq \mathbf{B}_{p,i}^{(l)}({\mathbf{H}_i^\mathrm{DD}}^{(l)})$ all correspond to the $l$-th realization in ${\mathbb{H}_i^\mathrm{DD}}^{(L)}$. For notational convenience, we define $\mathbf{A} \triangleq \{\mathbf{A}_{c,i}^\mathcal{L},\mathbf{A}_{p,i}^\mathcal{L} \mid i \in \mathcal{I}\}$, where $\mathbf{A}_{c,i}^\mathcal{L} \triangleq \{\mathbf{A}_{c,i}^{(l)} \mid l \in \mathcal{L}\}$ and $\mathbf{A}_{p,i}^\mathcal{L} \triangleq \{\mathbf{A}_{p,i}^{(l)} \mid l \in \mathcal{L}\}$. Similarly, we define $\mathbf{B} \triangleq \{\mathbf{B}_{c,i}^\mathcal{L},\mathbf{B}_{p,i}^\mathcal{L} \mid i \in \mathcal{I}\}$, where $\mathbf{B}_{c,i}^\mathcal{L} \triangleq \{\mathbf{B}_{c,i}^{(l)} \mid l \in \mathcal{L}\}$ and $\mathbf{B}_{p,i}^\mathcal{L} \triangleq \{\mathbf{B}_{p,i}^{(l)} \mid l \in \mathcal{L}\}$. Adopting the same approach as that utilized in proving~\eqref{eq:rate-wmmse}, the average rate\textendash\gls{wmmse} can be written as
\begin{subequations}
    \label{eq:ar-wmmse-sampled}
    \begin{align}
        \left(\hat{{\xi}}_{c,i}^{\mathrm{MMSE}}\right)^{(L)}
         & \triangleq
        \min_{\mathbf{A}_{c,i}, \mathbf{B}_{c,i}}
        \hat{{\xi}}_{c,i}^{(L)}
        = MN - \hat{R}_{c,i}^{(L)}, \\
        \left(\hat{{\xi}}_{p,i}^{\mathrm{MMSE}}\right)^{(L)}
         & \triangleq
        \min_{\mathbf{A}_{p,i}, \mathbf{B}_{p,i}}
        \hat{{\xi}}_{p,i}^{(L)}
        = MN - \hat{R}_{p,i}^{(L)}.
    \end{align}
\end{subequations}
The set of optimal \gls{mmse} equalizer corresponding to~\eqref{eq:ar-wmmse-sampled} are expressed as
$\mathbf{A}_{c,i}^{\mathrm{MMSE}} \triangleq \bigl\{ \mathbf{A}_{c,i}^{\mathrm{MMSE}(l)} \,\big|\, l \in \mathcal{L} \bigr\},\,
    \mathbf{A}_{p,i}^{\mathrm{MMSE}} \triangleq \bigl\{ \mathbf{A}_{p,i}^{\mathrm{MMSE}(l)} \,\big|\, l \in \mathcal{L} \bigr\}$.
Similarly, the set of corresponding optimal \gls{mmse} weight matrices are defined as
$\mathbf{B}_{c,i}^{\mathrm{MMSE}} \triangleq \bigl\{ \mathbf{B}_{c,i}^{\mathrm{MMSE}(l)} \,\big|\, l \in \mathcal{L} \bigr\}, \,
    \mathbf{B}_{p,i}^{\mathrm{MMSE}} \triangleq \bigl\{ \mathbf{B}_{p,i}^{\mathrm{MMSE}(l)} \,\big|\, l \in \mathcal{L} \bigr\}$.
By aggregating all $ I $ users, the \gls{mmse} equalizer and weight collections are given by
$\mathbf{A}^{\mathrm{MMSE}} \triangleq \bigl\{ \mathbf{A}_{c,i}^{\mathrm{MMSE}}, \mathbf{A}_{p,i}^{\mathrm{MMSE}} \,\big|\, i \in \mathcal{I} \bigr\},
    \,
    \mathbf{B}^{\mathrm{MMSE}} \triangleq \bigl\{ \mathbf{B}_{c,i}^{\mathrm{MMSE}}, \mathbf{B}_{p,i}^{\mathrm{MMSE}} \,\big|\, i \in \mathcal{I} \bigr\}$.
Building upon the relationship established in~\eqref{eq:ar-wmmse-sampled}, the deterministic average \gls{awmse} minimization problem is formulated as
\begin{subequations}
    \label{prob:aawmse}
    \begin{align}
        \mathcal{P}_{3}:\, \min_{\substack{\mathbf{P}_\mathrm{BF}, \boldsymbol{\Psi},                                  \\ \hat{\boldsymbol{\mu}},\mathbf{A}, \mathbf{B}}}
         & \quad \sum_{i=1}^{I} \hat{\xi}_{tot,i}^{(L)}\label{prob:aawmse_obj}                                         \\
        \text{s.t.} \quad
         & \quad  \sum_{i=1}^{I}\hat{\mu}_i + MN \geq \hat{\xi}_{c,i}^{(L)},\, i\in\mathcal{I} \label{prob:aawmse_pri} \\
         & \quad \hat{\boldsymbol{\mu}} \leq \mathbf{0}, \label{prob:aawmse_c}                                         \\
         & \quad \eqref{prob:asr_c},\,\eqref{prob:asr_d},
    \end{align}
\end{subequations}
where $\hat{\xi}_{\mathrm{tot},i}^{(L)} = \hat{\mu}_i + \hat{\xi}_{p,i}^{(L)} $ and $ \hat{\boldsymbol{\mu}} =\left[\hat{\mu}_1,\hat{\mu}_2, \ldots, \hat{\mu}_I\right] = -\hat{\mathbf{c}}$.
The optimization in~\eqref{prob:aawmse} with respect to $ (\mathbf{A}, \mathbf{B}) $ is obtained by minimizing the individual average \gls{awmse} in~\eqref{eq:ar-wmmse-sampled}, as the average \gls{awmse}s are decoupled with respect to their corresponding equalizers and weights.
This can be verified by showing that the \gls{kkt} optimality conditions of~\eqref{prob:aawmse} are satisfied for a given precoder $ \mathbf{P}_\mathrm{BF} $ and arrangement set $ \boldsymbol{\Psi} $ when $ \{\mathbf{A}, \mathbf{B}\} = \{\mathbf{A}^{\mathrm{MMSE}}, \mathbf{B}^{\mathrm{MMSE}}\} $.
Under this setting, problem~\eqref{prob:aawmse} simplifies to~\eqref{prob:saa}, demonstrating that the rate\textendash\gls{wmmse} equivalence holds not only at the global optimum but also at all stationary points.
Furthermore, any feasible points $ \{\mathbf{A}^*, \mathbf{B}^*, \mathbf{P}_\mathrm{BF}^*, \boldsymbol{\Psi}^*, \hat{\boldsymbol{\mu}}^*\} $ satisfying the \gls{kkt} conditions of~\eqref{prob:aawmse} also satisfies those of~\eqref{prob:saa} with $ \hat{\boldsymbol{\mu}} = -\hat{\mathbf{c}} $~\cite{7555358}.
Consequently, as $ L \to \infty $, the optimal solution of~\eqref{prob:aawmse} converges to that of~\eqref{prob:saa}, which in turn corresponds to the solution of the \gls{asr} problem in~\eqref{prob:asr}.

\subsection{Vectorization}
\par Unlike the \gls{siso} scenario considered in~\cite{10804646,10462183,10565898}, the matrix variables in~\eqref{prob:aawmse} makes the optimization process challenging.
Specifically, solving \eqref{prob:aawmse} requires managing $\hat{\xi}_{tot,i}^{(L)}$ in \eqref{prob:aawmse_obj} and $\hat{\xi}_{c,i}^{(L)}$ in \eqref{prob:aawmse_pri}, which are linear combinations with the instantaneous \gls{awmse} expressions ${\xi}_{tot,i}^{(l)}$ and ${\xi}_{c,i}^{(l)}$ in~\eqref{eq:awmse_opt_mse}. Unfortunately, the optimal solution of~\eqref{eq:awmse_opt_mse}, based on~\eqref{eq:opt_weight}, necessitates calculating the matrix inversions, $ \mathbf{K}_{c,i}^{-1} $ and $ \mathbf{K}_{p,i}^{-1} $, as shown in ~\eqref{eq:opt_mmse}. This ultimately introduces significant computational complexity and hinders the problem's convexity.

To avoid matrix inversion, we do not optimize the precoder directly using \eqref{eq:opt_weight}. Instead, we revert to the instantaneous \gls{awmse} expressions ${\xi}_{tot,i}^{(l)}$ and ${\xi}_{c,i}^{(l)}$ in \eqref{eq:awmse_opt_mse}, and expand its main components $ \tr(\mathbf{B}_{c,i}\mathbf{E}_{c,i}) $ as in \eqref{eq:vertor_expansion_c} at the top of next page.
After algebraic simplification, we obtain the whole expression of ${\xi}_{tot,i}^{(l)}$ and ${\xi}_{c,i}^{(l)}$  from \eqref{eq:awmse_opt_mse} in~\eqref{eq:vec_obj_common}, where $\tilde{\mathbf{p}}_c\triangleq\mathrm{vec}(\tilde{\mathbf{P}}_c)$, and $\tilde{\mathbf{p}}_{p,i}\triangleq\mathrm{vec}(\tilde{\mathbf{P}}_{p,i})$ for $i \in \mathcal{I}$.
Similarly, the corresponding formulation for $ \tr(\mathbf{B}_{p,i}\mathbf{E}_{p,i})-\log \det(\mathbf{B}_{p,i})$ from \eqref{eq:awmse_opt_mse} is derived in~\eqref{eq:vec_obj_private}.

\begin{figure*}[t!]
    \begin{equation}
        \label{eq:vertor_expansion_c}
        \begin{aligned}
            \tr\left\{\mathbf{B}_{c,i}\mathbf{E}_{c,i}\right\}
             & = \tr\left\{\mathbf{B}_{c,i} \mathbb{E}\left[ \left(\mathbf{A}_{c,i}\mathbf{y}_i^\mathrm{DD} - \mathbf{s}_c\right)\left(\mathbf{A}_{c,i}\mathbf{y}_i^\mathrm{DD} - \mathbf{s}_c\right)^H \right]\right\}              \\
             & = \tr\Big\{  \mathbf{B}_{c,i}\mathbf{A}_{c,i} \tilde{\mathbf{H}}_i^\mathrm{DD} \tilde{\mathbf{P}}_c \tilde{\mathbf{P}}_c^H {(\tilde{\mathbf{H}}_i^\mathrm{DD})}^H \mathbf{A}_{c,i}^H
            + \mathbf{B}_{c,i}\mathbf{A}_{c,i}  \sum_{i^\prime=1}^{I} {\tilde{\mathbf{H}}_i^\mathrm{DD}} \tilde{\mathbf{P}}_{p,i^\prime} \tilde{\mathbf{P}}_{p,i^\prime}^H {(\tilde{\mathbf{H}}_i^\mathrm{DD})}^H \mathbf{A}_{c,i}^H \\
             & \quad
            - \mathbf{B}_{c,i}\mathbf{A}_{c,i} {\tilde{\mathbf{H}}_i^\mathrm{DD}} \tilde{\mathbf{P}}_c
            - \mathbf{B}_{c,i}\tilde{\mathbf{P}}_c^H {(\tilde{\mathbf{H}}_i^\mathrm{DD})}^H \mathbf{A}_{c,i}^H
            + \sigma_{n,i}^2 \mathbf{B}_{c,i}\mathbf{A}_{c,i} \mathbf{A}_{c,i}^H + \mathbf{B}_{c,i}\mathbf{I}_{MN}  \Big\}.
        \end{aligned}
    \end{equation}
    \hrulefill
    \begin{subequations}
        \label{eq:vec_obj_all}
        \begin{align}
             & \tr(\mathbf{B}_{c,i}\mathbf{E}_{c,i})- \log \det(\mathbf{B}_{c,i})
            = \tilde{\mathbf{p}}_c^H \left[\mathbf{I}_{MN} \otimes {(\tilde{\mathbf{H}}_i^\mathrm{DD})}^H \mathbf{A}_{c,i}^H\mathbf{B}_{c,i}\mathbf{A}_{c,i}\tilde{\mathbf{H}}_i^\mathrm{DD} \right] \tilde{\mathbf{p}}_c
            + \sum_{i^\prime=1}^{I} \tilde{\mathbf{p}}_{p,i^\prime}^H \left[ \mathbf{I}_{MN} \otimes {(\tilde{\mathbf{H}}_i^\mathrm{DD})}^H \mathbf{A}_{c,i}^H\mathbf{B}_{c,i}\mathbf{A}_{c,i}\tilde{\mathbf{H}}_i^\mathrm{DD} \right] \tilde{\mathbf{p}}_{p,i^\prime} \notag  \\
             & \quad- \underbrace{\left[ \mathrm{vec}\left( {(\tilde{\mathbf{H}}_i^\mathrm{DD})}^H \mathbf{A}_{c,i}^H \mathbf{B}_{c,i}^H   \right) \right]^H \tilde{\mathbf{p}}_c}_{t_3}
            - \underbrace{\tilde{\mathbf{p}}_c^H \mathrm{vec}\left( {(\tilde{\mathbf{H}}_i^\mathrm{DD})}^H \mathbf{A}_{c,i}^H \mathbf{B}_{c,i}  \right)}_{t_4}
            + \sigma_{n,i}^2 \tr\left(\mathbf{B}_{c,i}\mathbf{A}_{c,i} \mathbf{A}_{c,i}^H \right) + \tr\left(\mathbf{B}_{c,i}\right)
            - \log \det(\mathbf{B}_{c,i}),\label{eq:vec_obj_common}                                                                                                                                                                                                            \\
             & \tr(\mathbf{B}_{p,i}\mathbf{E}_{p,i})- \log \det(\mathbf{B}_{p,i})
            = \tilde{\mathbf{p}}_{p,i}^H \left[ \mathbf{I}_{MN} \otimes {(\tilde{\mathbf{H}}_i^\mathrm{DD})}^H \mathbf{A}_{p,i}^H\mathbf{B}_{p,i}\mathbf{A}_{p,i}\tilde{\mathbf{H}}_i^\mathrm{DD} \right] \tilde{\mathbf{p}}_{p,i}
            + \sum_{i^\prime \neq i} \tilde{\mathbf{p}}_{p,i^\prime}^H \left[ \mathbf{I}_{MN} \otimes {(\tilde{\mathbf{H}}_i^\mathrm{DD})}^H \mathbf{A}_{p,i}^H\mathbf{B}_{p,i}\mathbf{A}_{p,i}\tilde{\mathbf{H}}_i^\mathrm{DD} \right] \tilde{\mathbf{p}}_{p,i^\prime} \notag \\
             & \quad - \left[ \mathrm{vec}\left( {(\tilde{\mathbf{H}}_i^\mathrm{DD})}^H \mathbf{A}_{p,i}^H \mathbf{B}_{p,i}^H \right) \right]^H \tilde{\mathbf{p}}_{p,i}
            - \tilde{\mathbf{p}}_{p,i}^H  \mathrm{vec}\left( {(\tilde{\mathbf{H}}_i^\mathrm{DD})}^H \mathbf{A}_{p,i}^H \mathbf{B}_{p,i}  \right)
            + \sigma_{n,i}^2 \tr\left(\mathbf{B}_{p,i}\mathbf{A}_{p,i} \mathbf{A}_{p,i}^H \right) + \tr\left(\mathbf{B}_{p,i}\right)
            - \log \det(\mathbf{B}_{p,i}),\label{eq:vec_obj_private}
        \end{align}
    \end{subequations}
    \hrulefill
\end{figure*}

\par However, the expressions in~\eqref{eq:vec_obj_all} are cumbersome. Closer inspection reveals that the third and fourth terms, $t_3$ and $t_4$, in~\eqref{eq:vec_obj_common} are inherently connected, and the same observation holds for the corresponding terms in~\eqref{eq:vec_obj_private}.
We further establish that the weight matrices $\mathbf{B}_{c,i}$ and $\mathbf{B}_{p,i}$ are Hermitian when taking the optimal values, that is, $\mathbf{B}_{c,i}^{\mathrm{MMSE}} = (\mathbf{B}_{c,i}^{\mathrm{MMSE}})^H$ and $\mathbf{B}_{p,i}^{\mathrm{MMSE}} = (\mathbf{B}_{p,i}^{\mathrm{MMSE}})^H$. 
These properties enable \eqref{eq:vec_obj_common} and \eqref{eq:vec_obj_private} to be simplified and expressed in compact forms, as shown in~\eqref{eq:vec_obj_common_simplify} and~\eqref{eq:vec_obj_private_simplify} at the top of next page, where
\begin{equation*}
    \begin{aligned}
        \mathbf{C}_{c,i} & = \mathbf{I}_{MN}  \otimes {(\tilde{\mathbf{H}}_i^\mathrm{DD})}^H \mathbf{A}_{c,i}^H\mathbf{B}_{c,i}\mathbf{A}_{c,i}\tilde{\mathbf{H}}_i^\mathrm{DD}, \\
        \mathbf{C}_{p,i} & = \mathbf{I}_{MN}  \otimes {(\tilde{\mathbf{H}}_i^\mathrm{DD})}^H \mathbf{A}_{p,i}^H\mathbf{B}_{p,i}\mathbf{A}_{p,i}\tilde{\mathbf{H}}_i^\mathrm{DD}, \\
        \mathbf{d}_{c,i} & = \mathrm{vec}( {(\tilde{\mathbf{H}}_i^\mathrm{DD})}^H \mathbf{A}_{c,i}^H \mathbf{B}_{c,i} ),\,
        \mathbf{d}_{p,i} = \mathrm{vec}( {(\tilde{\mathbf{H}}_i^\mathrm{DD})}^H \mathbf{A}_{p,i}^H \mathbf{B}_{p,i} ),                                                           \\
        d_{c,i}          & =\sigma_{n,i}^2 \tr(\mathbf{B}_{c,i}\mathbf{A}_{c,i} \mathbf{A}_{c,i}^H ) + \tr(\mathbf{B}_{c,i}) - \log \det(\mathbf{B}_{c,i}),                      \\
        d_{p,i}          & =\sigma_{n,i}^2 \tr(\mathbf{B}_{p,i}\mathbf{A}_{p,i} \mathbf{A}_{p,i}^H ) + \tr\left(\mathbf{B}_{p,i}\right) - \log \det(\mathbf{B}_{p,i}).
    \end{aligned}
\end{equation*}
\begin{figure*}
    \begin{subequations}
        \begin{align}
            \tr(\mathbf{B}_{c,i}\mathbf{E}_{c,i}) - \log \det(\mathbf{B}_{c,i})
             & = \tilde{\mathbf{p}}_c^H \mathbf{C}_{c,i} \tilde{\mathbf{p}}_c
            + \sum_{i^\prime=1}^{I} \tilde{\mathbf{p}}_{p,i^\prime}^H \mathbf{C}_{c,i} \tilde{\mathbf{p}}_{p,i^\prime}  - \mathbf{d}_{c,i}^H \tilde{\mathbf{p}}_c
            - \tilde{\mathbf{p}}_c^H \mathbf{d}_{c,i}
            + {d}_{c,i}\notag                                                         \\
             & = \tilde{\mathbf{p}}_c^H \mathbf{C}_{c,i} \tilde{\mathbf{p}}_c
            + \sum_{i^\prime=1}^{I} \tilde{\mathbf{p}}_{p,i^\prime}^H \mathbf{C}_{c,i} \tilde{\mathbf{p}}_{p,i^\prime}  - 2\mathrm{Re}\left\{\mathbf{d}_{c,i}^H \tilde{\mathbf{p}}_c\right\}
            + {d}_{c,i},         \label{eq:vec_obj_common_simplify}                   \\
            \tr(\mathbf{B}_{p,i}\mathbf{E}_{p,i}) - \log \det(\mathbf{B}_{p,i})
             & = \tilde{\mathbf{p}}_{p,i}^H \mathbf{C}_{p,i} \tilde{\mathbf{p}}_{p,i}
            + \sum_{i^\prime \neq i} \tilde{\mathbf{p}}_{p,i^\prime}^H \mathbf{C}_{p,i} \tilde{\mathbf{p}}_{p,i^\prime} - 2\mathrm{Re}\left\{\mathbf{d}_{p,i}^H \tilde{\mathbf{p}}_{p,i}\right\}
            + {d}_{p,i}.        \label{eq:vec_obj_private_simplify}
        \end{align}
    \end{subequations}    \hrulefill
\end{figure*}

\subsection{Alternating Optimization}
\par Problem~\eqref{prob:aawmse} is inherently non-convex when jointly optimizing the variables $ \mathbf{A} $, $ \mathbf{B} $, $ \mathbf{P}_\mathrm{BF} $, $\boldsymbol{\Psi}$ and $ \hat{\boldsymbol{\mu}} $.
However, it becomes convex with respect to each variable block when the remaining ones are fixed.
Therefore, an \gls{ao} strategy is adopted, consisting of three main steps:
(1) The equalizers $\mathbf{A}$ and weights $\mathbf{B}$ are updated while keeping the precoder $\mathbf{P}_{\mathrm{BF}}$ and the arrangement matrix set $\boldsymbol{\Psi}$ fixed.
(2) The precoder $\mathbf{P}_{\mathrm{BF}}$ and message split $\hat{\boldsymbol{\mu}}$ are optimized for the fixed $\boldsymbol{\Psi}$ and the updated $\mathbf{A}$ and $\mathbf{B}$.
(3) The arrangement matrix set $\boldsymbol{\Psi}$ is then updated based on the updated $\mathbf{P}_{\mathrm{BF}}$, $\mathbf{A}$, and $\mathbf{B}$.
\begin{enumerate}
    \item \textit{Update equalizers and weights}. For each channel realization, in the $k$-th iteration of the \gls{ao} algorithm, the equalizers and weights are updated according to
          \begin{equation}
              (\mathbf{A}, \mathbf{B}) =
              \big( \mathbf{A}^{\mathrm{MMSE}}(\mathbf{X}^{[k-1]}),\, \mathbf{B}^{\mathrm{MMSE}}(\mathbf{X}^{[k-1]}) \big),
          \end{equation}
          where  $\mathbf{X}^{[k-1]} \triangleq (\mathbf{P}_\mathrm{BF}^{[k-1]}, \boldsymbol{\Psi}^{[k-1]})$, therein, $\mathbf{P}_\mathrm{BF}^{[k-1]}$ and $\boldsymbol{\Psi}^{[k-1]}$ are the precoding and arrangement matrices obtained from the $(k-1)$-th iteration. To facilitate the subsequent optimization of $\mathbf{P}_\mathrm{BF}$, we define a set of \gls{saf} as
          $\{\hat{\mathbf{C}}_{c,i},\hat{\mathbf{C}}_{p,i},\hat{\mathbf{d}}_{c,i},\hat{\mathbf{d}}_{p,i},\hat{d}_{c,i},\hat{d}_{p,i}\}$,
          which are computed based on the updated $(\mathbf{A}, \mathbf{B})$.
          Specifically, they are defined as the ensemble means of their instantaneous values:
          \begin{align*}
              \hat{\mathbf{C}}_{c,i} & = \frac{1}{L}\sum_{l=1}^{L} \mathbf{C}_{c,i}^{(l)},  \,
              \hat{\mathbf{d}}_{c,i}  = \frac{1}{L}\sum_{l=1}^{L} \mathbf{d}_{c,i}^{(l)},  \,
              \hat{d}_{c,i}           = \frac{1}{L}\sum_{l=1}^{L} d_{c,i}^{(l)},               \\
              \hat{\mathbf{C}}_{p,i} & = \frac{1}{L}\sum_{l=1}^{L} \mathbf{C}_{p,i}^{(l)},  \,
              \hat{\mathbf{d}}_{p,i}  = \frac{1}{L}\sum_{l=1}^{L} \mathbf{d}_{p,i}^{(l)},  \,
              \hat{d}_{p,i}           = \frac{1}{L}\sum_{l=1}^{L} d_{p,i}^{(l)}.
          \end{align*}
    \item \textit{Update precoders}. Given the arrangement matrix set and the updated equalizers and weights, the precoder update problem is formulated by substituting $\mathbf{A}^{\mathrm{MMSE}}(\mathbf{X}^{[k-1]})$, $\mathbf{B}^{\mathrm{MMSE}}(\mathbf{X}^{[k-1]})$ and \gls{saf} set of dependent entities into~\eqref{prob:aawmse}, leading to
          \begin{subequations}
              \label{prob:aawmse_simplify}
              \begin{align}
                  \mathcal{P}_{4}:\, \min_{\mathbf{P}_\mathrm{BF}, \hat{\boldsymbol{\mu}}}
                   & \quad t                                                                                                                         \\
                  \text{s.t.} \quad
                   & \quad \sum_{i=1}^{I}\Big(\hat{\mu}_i + \hat{\phi}_{p,i}\Big)-t\leq 0, \label{prob:aawmse_simplify_b}                            \\
                   & \quad  \hat{\phi}_{c,i}-\sum_{i^\prime=1}^{I}\hat{\mu}_{i^\prime} - MN \leq 0,\, i\in\mathcal{I} \label{prob:aawmse_simplify_c} \\
                   & \quad \eqref{prob:asr_c},\,\eqref{prob:asr_d},\,\eqref{prob:aawmse_c},
              \end{align}
          \end{subequations}
          where $\hat{\phi}_{c,i} = \tilde{\mathbf{p}}_c^H \hat{\mathbf{C}}_{c,i} \tilde{\mathbf{p}}_c
              + \sum_{i^\prime=1}^{I} \tilde{\mathbf{p}}_{p,i^\prime}^H \hat{\mathbf{C}}_{c,i} \tilde{\mathbf{p}}_{p,i^\prime}
              - 2\mathrm{Re}\left\{\hat{\mathbf{d}}_{c,i}^H \tilde{\mathbf{p}}_c\right\}
              + \hat{d}_{c,i}$ and $\hat{\phi}_{p,i} = \tilde{\mathbf{p}}_{p,i}^H \hat{\mathbf{C}}_{p,i} \tilde{\mathbf{p}}_{p,i}
              + \sum_{i^\prime \neq i} \tilde{\mathbf{p}}_{p,i^\prime}^H \hat{\mathbf{C}}_{p,i} \tilde{\mathbf{p}}_{p,i^\prime}
              - 2\mathrm{Re}\left\{\hat{\mathbf{d}}_{p,i}^H \tilde{\mathbf{p}}_{p,i}\right\}
              + \hat{d}_{p,i}$.
          Problem~\eqref{prob:aawmse_simplify} constitutes a convex \gls{qcqp}, which can be efficiently solved via interior-point methods~\cite{boyd2004convex,ye2011interior}.
    \item \textit{Update arrangement matrices}. Given the updated equalizers, weights, and precoder, the arrangement matrices update problem is formulated by substituting $\mathbf{P}_\mathrm{BF}^{[k]}$ into~\eqref{prob:aawmse_simplify}, leading to
          \begin{subequations}
              \label{prob:aawmse_update_psi}
              \begin{align}
                  \mathcal{P}_{5}:\, \min_{\boldsymbol{\Psi}}
                   & \quad t                                                                                                                                   \\
                  \text{s.t.} \quad
                   & \quad \eqref{prob:aawmse_simplify_b},\,\eqref{prob:aawmse_simplify_c},\, \eqref{prob:asr_c},\,\eqref{prob:asr_d},\,\eqref{prob:aawmse_c}.
              \end{align}
          \end{subequations}
          Problem~\eqref{prob:aawmse_update_psi} constitutes a mixed-integer problem, which can be solved via MOSEK optimization toolbox~\cite{aps2019mosek}.
\end{enumerate}
The three steps are alternately executed until convergence, as summarized in Algorithm~\ref{algo:ao}, where $\epsilon$ denotes an arbitrarily small threshold. The problems \eqref{prob:aawmse_simplify} and \eqref{prob:aawmse_update_psi} are solved by CVX toolbox in Matlab \cite{grant2008cvx,aps2019mosek}.
For a given sample set ${\mathbb{H}_i^{\mathrm{DD}}}^{(L)}$, the sequence of optimized variables generated by Algorithm~\ref{algo:ao} converges to the set of \gls{kkt} points of the corresponding sampled \gls{asr} problem in~\eqref{prob:saa}.
Furthermore, as $L \to \infty$, the iterates converge almost surely to the set of \gls{kkt} solutions of the original \gls{asr} problem in~\eqref{prob:asr} \cite{7555358}.

\begin{algorithm}[t!]
    \caption{Alternating Optimization algorithm}\label{algo:ao}
    \textbf{Initialize}: $k=0$, $\mathbf{P}_\mathrm{BF}^{[0]}$\;
    \Repeat{$t^{[k]}-t^{[k-1]}<\epsilon$}{
    {$k\leftarrow k+1$;\, $\mathbf{A}=\mathbf{A}^{\mathrm{MMSE}}(\mathbf{P}_\mathrm{BF}^{[k-1]})$,\,$\mathbf{B}=\mathbf{B}^{\mathrm{MMSE}}(\mathbf{P}_\mathrm{BF}^{[k-1]})$}\;
    {Compute $\hat{\mathbf{C}}_{c,i}$, $\hat{\mathbf{C}}_{p,i}$, $\hat{\mathbf{d}}_{c,i}$, $\hat{\mathbf{d}}_{p,i}$, $\hat{d}_{c,i}$, $\hat{d}_{p,i}$, $\forall i\in\mathcal{I}$}\;
    {Solve problem \eqref{prob:aawmse_simplify} and denote the optimal solutions as $\mathbf{P}_\mathrm{BF}^{*}$, $\hat{\boldsymbol{\mu}}^{*}$}\;
    {Solve problem \eqref{prob:aawmse_update_psi} and denote the optimal value of the objective function as $t^{*}$ and the optimal solutions as $\boldsymbol{\Psi}^{*}$}\;
    Update $t^{[k]}\leftarrow t^*$, $\mathbf{P}_\mathrm{BF}^{[k]}\leftarrow\mathbf{P}_\mathrm{BF}^{*}$, $\hat{\boldsymbol{\mu}}^{[k]}\leftarrow\hat{\boldsymbol{\mu}}^{*}$, $\boldsymbol{\Psi}^{[k]}\leftarrow\boldsymbol{\Psi}^{*}$\;
    }
\end{algorithm}
\section{Simulation Results}\label{sec:results}
\par In this section, we evaluate the performance of the proposed \gls{rsma}-\gls{otfs} scheme. The performance of the proposed scheme is compared with the following baseline schemes.
    {SDMA-OTFS}: A conventional \gls{sdma} scheme where each user's data is transmitted independently using OTFS modulation, i.e., no power is allocated to common streams.
    {NOMA-OTFS}: In a $I$-user \gls{noma} scheme, $i$-th user message is decoded after $i-1$ user's signal are removed. We assume that the user with higher channel gain decodes and removes the interference from the user with lower channel gain before decoding its own signal. Specifically, for $2$-user case, \gls{noma} is a particular instance of \gls{rsma}, i.e., no power is allocated to one private stream and the entire common stream is utilized to transmit the message of a single user \cite{8907421,Mao2018}.
The simulation setup is summarized in Table~\ref{tab:simu_parameters}. The performance metric used in the simulations is the \gls{esr}, which is averaged over 100 random channel realizations.
\begin{table}[t!]
    \caption{Simulation Parameters}
    \label{tab:simu_parameters}
    \begin{tabular}{IcIcI}
        \bottomrule[1.05pt]
        \hline
        Parameters                       & \multicolumn{1}{cI}{Values}                       \\ \toprule[1.05pt]
        \bottomrule[1.05pt]
        \gls{leo} satellite altitude     & $500$ km                                          \\ \hline
        \gls{leo} satellite velocity     & $7.58\times10^3$ m/s                              \\ \hline
        Carrier frequency                & $7.6$ GHz                                         \\ \hline
        Subcarrier spacing               & $480$ kHz                                         \\ \hline
        Bandwidth                        & $128.88$ MHz                                      \\ \hline
        Number of \gls{nlos} paths       & 2                                                 \\ \hline
        Rician factor                    & $10$ dB                                           \\ \hline
        \gls{aod} range                  & $ \theta_{i,q}\in[-\frac{\pi}{2},\frac{\pi}{2}) $ \\ \hline
        \gls{nlos} channel gain variance & $ \sigma_{i,q}^2=1 $                              \\ \hline
        \gls{snr} of the pilot signal    & $\gamma_e=30\,\mathrm{dB}$                        \\ \hline
        Channel error samples            & $ L=1000 $                                        \\ \hline
        \toprule[1.05pt]
    \end{tabular}
\end{table}
\par
\begin{figure}[t!]
    \centering
    \includegraphics[width=1.05\textwidth]{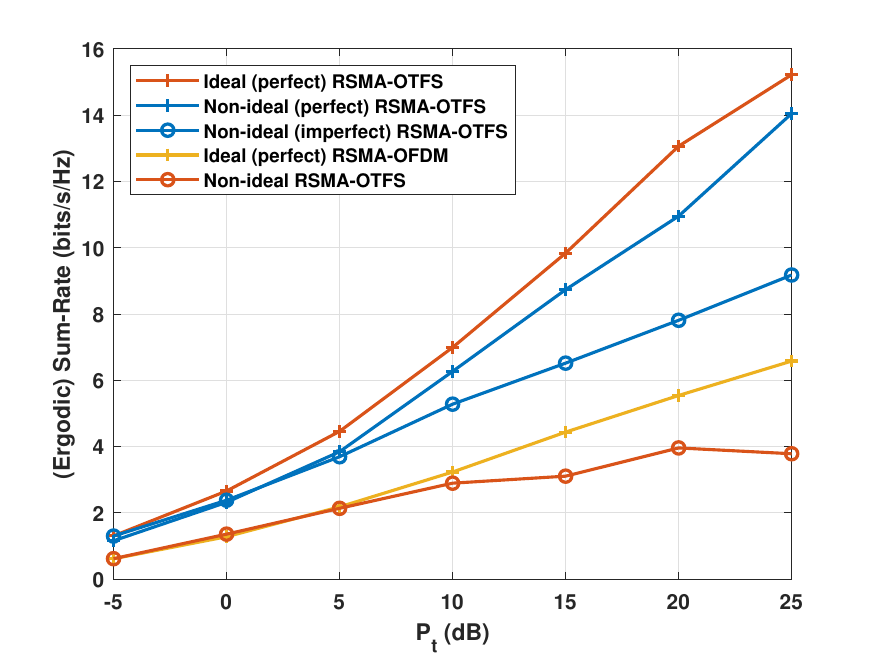}
    \caption{Ergodic sum-rate comparison of different strategies with/without fractional Delay, Doppler Shift and rectangular pulse shaping effects, $M=N=4$, $N_t=2$, $I=2$, $\rho=0.7$.}
    \label{fig:1-without_fractional}
\end{figure}

 First, Fig.~\ref{fig:1-without_fractional} illustrates the performance loss that arises with fractional delay and Doppler shifts, as well as the with the practical non-ideal rectangular pulse shaping. The curve labeled ``Ideal (perfect) RSMA-OTFS'' corresponds to a system model that does not incorporate these impairments, whereas ``Non-ideal RSMA-OTFS'' evaluates the performance under realistic conditions using the optimized precoders of the idealized model. A substantial degradation is observed when the impairments are omitted, and the loss becomes more pronounced in high \gls{snr} regimes. The curves ``Non-ideal (perfect) RSMA-OTFS'' and ``Non-ideal (imperfect) RSMA-OTFS'' represent the cases with perfect and imperfect \gls{csit}, respectively, under realistic conditions. The proposed \gls{rsma}-\gls{otfs} approach, under both perfect and imperfect \gls{csit}, achieves substantial performance gains over conventional \gls{rsma}-\gls{ofdm} modulation in high-mobility scenarios. Notably, the performance of ``Non-ideal RSMA-OTFS'' falls below that of ``Non-ideal (imperfect) RSMA-OTFS'', indicating that the combined impact of fractional delay, Doppler shifts, and rectangular pulse shaping is more severe than that of imperfect \gls{csit}. These observations demonstrate the necessity of evaluating and optimizing system performance under realistic conditions in \gls{otfs}-based systems.

\par
\begin{figure}[t!]
    \begin{subfigure}{.49\textwidth}
        \centering
        \hspace{-6mm} \includegraphics[scale=0.33]{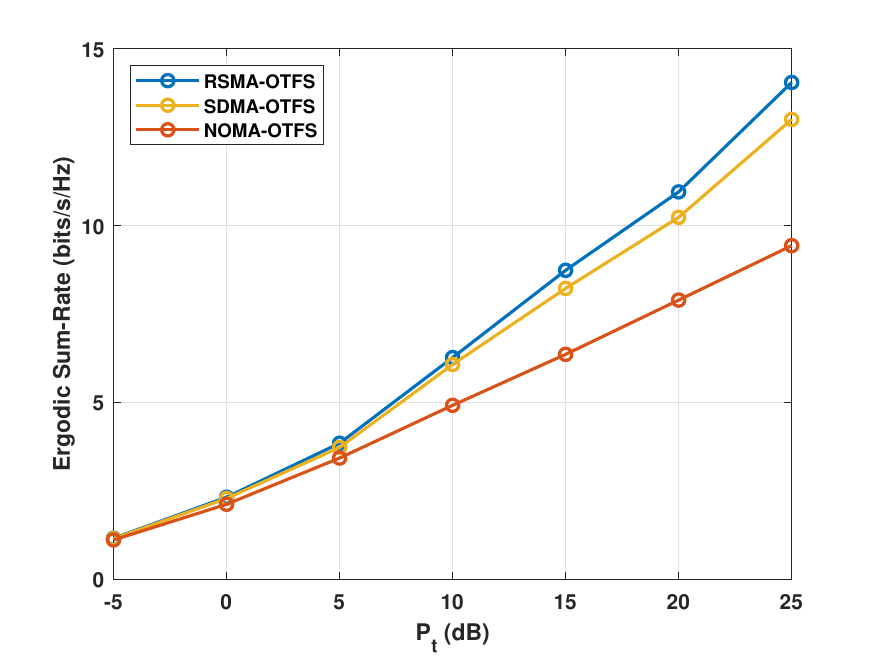}
        \caption{Perfect CSIT}
        \label{fig:CSIT_quality_perfect}
    \end{subfigure}
    \begin{subfigure}{.49\textwidth}
        \centering
        \hspace{-4mm} \includegraphics[scale=0.33]{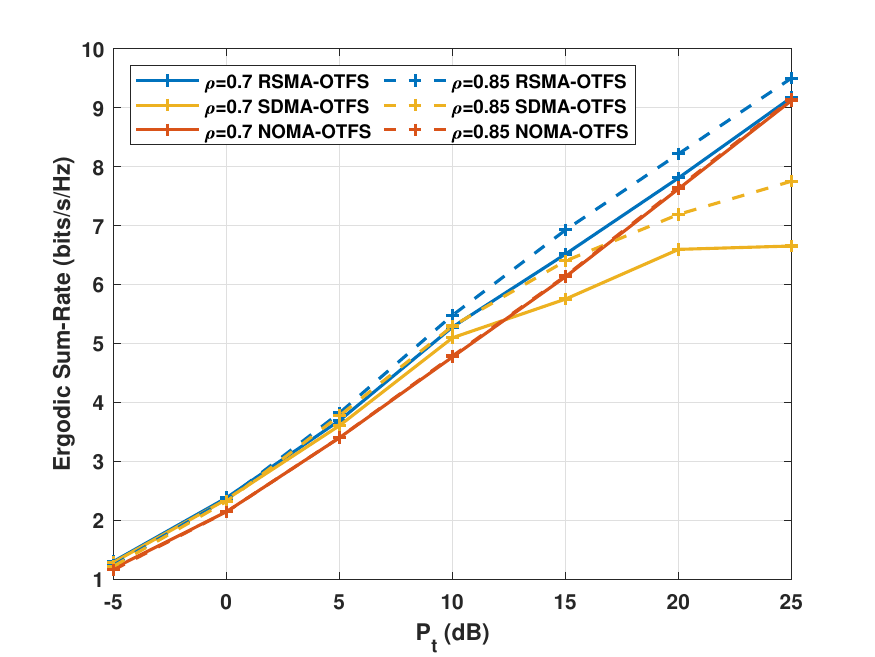}
        \caption{Imperfect CSIT}
        \label{fig:CSIT_quality_imperfect}
    \end{subfigure}
    \caption{Ergodic sum-rate comparison of different strategies under perfect and imperfect CSIT conditions, $M=N=4$, $N_t=2$, $I=2$.}
    \label{fig:per_imp_compare}
\end{figure}

 The impact of imperfect \gls{csit} is illustrated in Fig.~\ref{fig:CSIT_quality_perfect} and Fig.~\ref{fig:CSIT_quality_imperfect}. The results show that \gls{rsma}-\gls{otfs} achieves higher \gls{esr} than \gls{sdma}-\gls{otfs} and \gls{noma}-\gls{otfs} under both perfect and imperfect \gls{csit}. All schemes experience performance degradation when the \gls{csit} quality decreases, yet the reduction is more significant for the \gls{sdma}-\gls{otfs} scheme and
Fig.~\ref{fig:CSIT_quality_perfect} shows the \gls{dof} loss incurred by \gls{noma}-\gls{otfs} as predicted in \cite{9451194}.
This behavior confirms that \gls{rsma}-\gls{otfs} is intrinsically more resilient to \gls{csit} imperfections because the common stream provides additional flexibility for interference mitigation. Although \gls{noma}-\gls{otfs} exhibits a similar trend, the performance gap between perfect and imperfect \gls{csit} remains smaller, since \gls{noma} does not fully benefit from accurate \gls{csit}.
The influence of varying \gls{csit} quality is further examined in Fig.~\ref{fig:CSIT_quality_imperfect}. As the channel correlation coefficient $\rho$ decreases, indicating more severe \gls{csit} imperfections, all schemes experience a decline in \gls{esr}. The reduction stems from the limited accuracy in precoding and interference management when \gls{csit} deteriorates. Notably, the relative gain\footnote{The relative gain between two schemes $x$ and $y$ is defined as $(x-y)/x$.} of \gls{rsma}-\gls{otfs} over \gls{sdma}-\gls{otfs} increases from $12.49\%$ to $15.54\%$ when $\rho$ decreases from $0.85$ to $0.7$ at $P_\mathrm{t}=20$ dB, which confirms that \gls{rsma}-\gls{otfs} retains its advantage even as the \gls{csit} quality worsens. Across all \gls{csit} conditions, the proposed \gls{rsma}-\gls{otfs} scheme maintains superior performance, demonstrating its suitability for multi-user multi-antenna systems with imperfect or partial \gls{csit}.

\par
\begin{figure}[t!]
    \centering
    \includegraphics[width=1.05\textwidth]{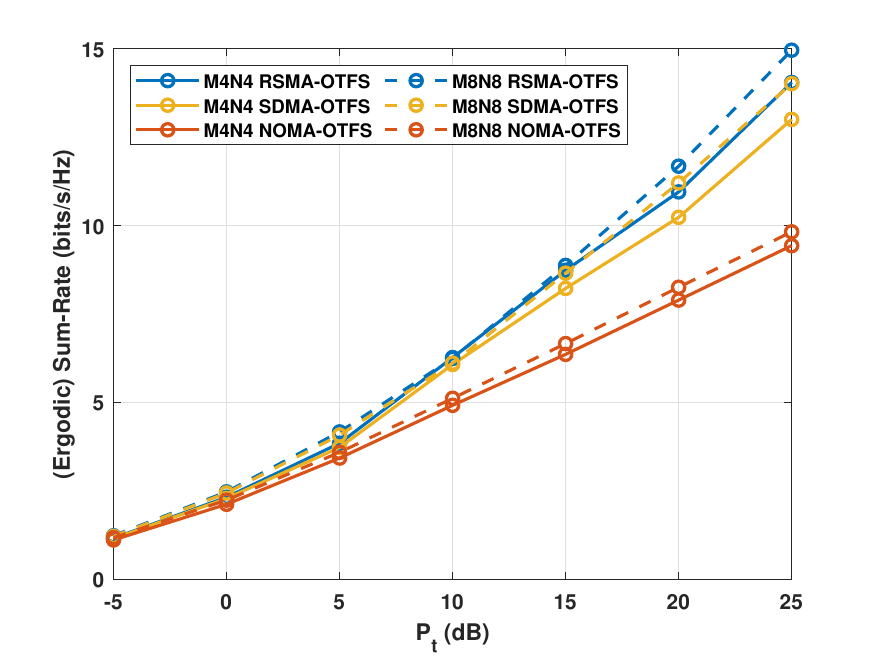}
    \caption{Sum-rate comparison of different strategies with different OTFS frame resources, $N_t=2$, $I=2$, $\rho=0.7$.}
    \label{fig:2-resources_compare}
\end{figure}

 Next, the impact of \gls{otfs} frame resources on the system performance is evaluated and shown in Fig.~\ref{fig:2-resources_compare}. The notations ``M4N4 RSMA-OTFS'', and ``M8N8 RSMA-OTFS'' represent the proposed \gls{rsma}-\gls{otfs} scheme with \gls{otfs} frame configurations of $M=4$, $N=4$ and $M=8$, $N=8$, respectively. Similar notations are used for the baseline schemes.
It can be observed that increasing the number of subcarriers $M$ and the number of time slots $N$ leads to performance improvement in the \gls{esr} performance for all schemes. This is because larger \gls{otfs} frame resources enhance the delay and Doppler resolution for channel estimation, hence reducing the effect of fractional delay and Doppler shift, which enhances the overall system performance. Additionally, it is evident that the proposed \gls{rsma}-\gls{otfs} scheme consistently outperforms the baseline schemes across different \gls{otfs} frame configurations, demonstrating its effectiveness in leveraging the available resources.

\par
\begin{figure}[t!]
    \centering
    \begin{subfigure}[b]{.49\textwidth}
        \centering
        \hspace{-6mm} \includegraphics[scale=0.33]{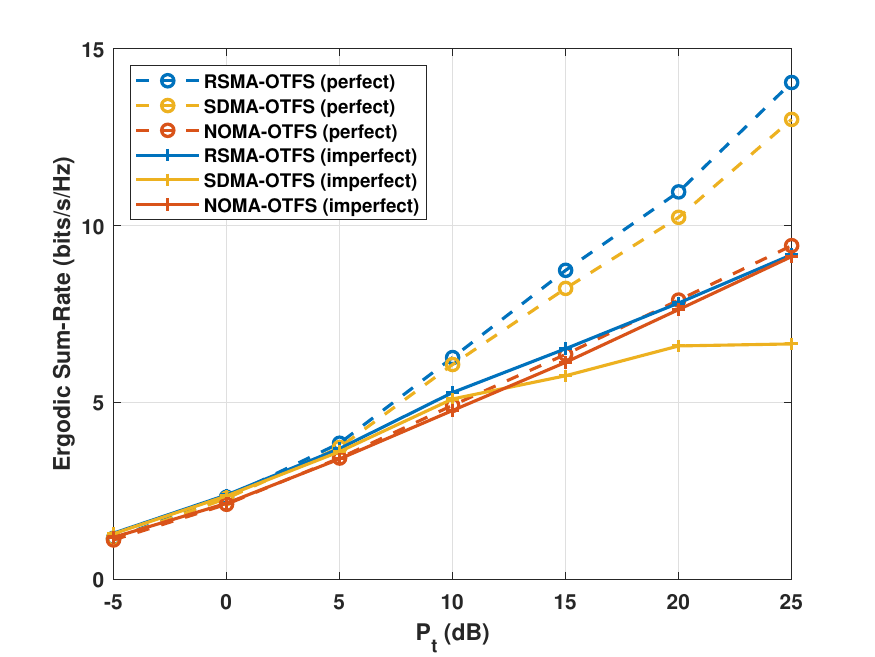}
        \caption{Underloaded, $I=2$}
        \label{fig:underloaded}
    \end{subfigure}
    \begin{subfigure}[b]{.49\textwidth}
        \centering
        \hspace{-4mm} \includegraphics[scale=0.33]{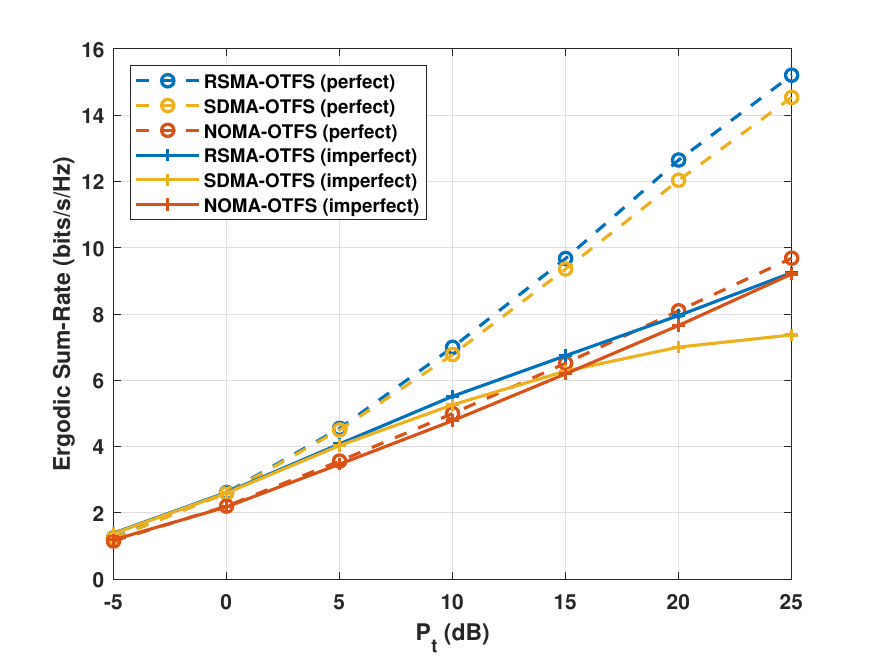}
        \caption{Overloaded, $I=4$}
        \label{fig:overloaded}
    \end{subfigure}
    \caption{Ergodic sum-rate comparison of different strategies with underloaded and overloaded deployments, $M=N=4$, $N_t=2$.}
    \label{fig:under_over_compare}
\end{figure}

 Fig.~\ref{fig:underloaded} and Fig.~\ref{fig:overloaded} present the performance comparison under underloaded and overloaded user deployments. In the underloaded setting, the number of transmit antennas is greater than or equal to the number of users, whereas in the overloaded setting the number of users exceeds the number of transmit antennas. Across both scenarios, the proposed \gls{rsma}-\gls{otfs} scheme achieves higher performance than the \gls{sdma}-\gls{otfs} and \gls{noma}-\gls{otfs} baselines.
The figures further show that when the system transitions from the underloaded case ($I=2$) to the overloaded case ($I=4$), the relative rate gain of \gls{rsma}-\gls{otfs} over \gls{sdma}-\gls{otfs} decreases from $6.56\%$ to $4.83\%$ under perfect \gls{csit}, and from $15.50\%$ to $11.97\%$ under imperfect \gls{csit}, with $P_\mathrm{t}=20$~dB. This indicates that \gls{rsma}-\gls{otfs} remains effective in both scenarios, and the gain is more evident when the transmitter has imperfect channel knowledge. As the number of users increases, the relative gain becomes smaller because both \gls{rsma} and \gls{sdma} may deactivate one private stream when the number of users exceeds the number of transmit antennas, which reduces the performance gap. In addition, the contribution of the common stream to the overall multiplexing gain becomes less influential compared with that of the private streams.
The results demonstrate that the proposed \gls{rsma}-\gls{otfs} scheme consistently outperforms the \gls{sdma}-\gls{otfs} and \gls{noma}-\gls{otfs} baselines across different user deployment configurations.

\section{Conclusion}\label{sec:conclusion}
\par This work investigated the integration of \gls{rsma} and \gls{otfs} for downlink multi-user multi-antenna transmission in \gls{leo} satellite systems, where rapid channel variations and imperfect \gls{csit} pose significant challenges. A robust transmission framework was developed to account for practical propagation effects, including rectangular pulse shaping, fractional Doppler shifts, fractional delays, and statistical \gls{csit}. Within this framework, a compact cross-domain input-output relationship was derived, enabling a tractable problem optimization.
Based on the proposed model, an \gls{esr} maximization problem was formulated by jointly optimizing the precoders, message splitting, power allocation, and precoder arrangement. The resulting non-convex problem was solved using a \gls{wmmse}-based \gls{ao} algorithm that does not rely on sparsity assumptions of the \gls{dd}-domain channel.
Numerical results showed that ignoring practical impairments can lead to severe performance degradation, while the proposed multi-antenna \gls{rsma}-\gls{otfs} scheme consistently achieves higher \gls{esr} and improved robustness to \gls{csit} uncertainty across various channel conditions and user deployments. These results highlight the effectiveness of the proposed framework and its potential for future \gls{leo} satellite communication systems.

\bibliographystyle{IEEEtran}
\bibliography{IEEEabrv,ref_OTFS_RSMA_only}


\end{document}